\documentclass[11pt]{article}
\usepackage{jheppub}

\usepackage{hyperref}

\usepackage[usenames,dvipsnames,table]{xcolor}
\usepackage{graphicx,amsmath,amssymb,multirow,array,bm,mathrsfs}
\usepackage{epsf,amsfonts}
\usepackage[numbers,sort&compress]{natbib}

\usepackage{slashed}
\usepackage[yyyymmdd,hhmmss]{datetime}

\usepackage{color}
\usepackage{hyperref}
\hypersetup{hidelinks}
\usepackage{leftidx}
\usepackage{subcaption}
\usepackage[utf8]{inputenc}
\usepackage[section]{placeins}
\usepackage{physics}

\usepackage{tikz}
\usetikzlibrary{decorations.markings}
\usetikzlibrary{decorations.pathmorphing}

\newcommand{\be}{\begin{equation}}
\newcommand{\ee}{\end{equation}}
\newcommand{\ba}{\begin{aligned}}
\newcommand{\ea}{\end{aligned}}

\newcommand{\cO}{\mathcal{O}}

\newcommand{\eps}{\varepsilon}

\newcommand{\legP}{\mathcal{P}}
\newcommand{\prob}{\rho}
\def\MM{\mathcal{M}}

\def\avg#1{\left\langle #1\right\rangle}
\def\Res{\operatorname*{Res}}

\usepackage{xcolor}
\usepackage{comment}

\title{Extremal Effective Field Theories}

\author[a]{Simon Caron-Huot,}
\emailAdd{schuot@physics.mcgill.ca}
\author[a]{Vincent Van Duong}
\emailAdd{vvd21@cam.ac.uk}

\affiliation[a]{Department of Physics, McGill University, Montr\'{e}al, QC, Canada}

\abstract{
Effective field theories (EFT) parameterize the long-distance effects of short-distance dynamics whose details may or may not be known.
Previous work showed that EFT coefficients must obey certain positivity constraints if causality
and unitarity are satisfied at all scales.
We explore those constraints from the perspective of $2\to 2$ scattering amplitudes of a light real scalar field,
using semi-definite programming to carve out the space of allowed EFT coefficients for a given mass threshold $M$.
We point out that all EFT parameters are bounded both below and above, effectively showing that dimensional analysis scaling is a consequence of causality.
This includes the coefficients of $s^2+t^2+u^2$ and $stu$ type interactions.
We present simple $2\to2$ extremal amplitudes which realize, or ``rule in'', kinks in coefficient space and whose convex hull span a large fraction
of the allowed space.
}

\begin{document}
\maketitle
\flushbottom

\def\be{\begin{equation}}
\def\ee{\end{equation}}
\newcommand{\lsim}{\mathrel{\hbox{\rlap{\lower.55ex \hbox{$\sim$}} \kern-.3em \raise.4ex \hbox{$<$}}}}
\newcommand{\gsim}{\mathrel{\hbox{\rlap{\lower.55ex \hbox{$\sim$}} \kern-.3em \raise.4ex \hbox{$>$}}}}
\renewcommand{\abstractname}{\vspace{-\baselineskip}}
\def\nl{\nonumber\\}
\def\ket#1{\big| #1\big\rangle}
\def\l{\langle}
\def\r{\rangle}
\def\MM{\mathcal{M}}
\def\AA{\mathcal{A}}
\def\phib{\bar{\phi}}
\def\eps{\epsilon}

\newmuskip\pFqmuskip
\newcommand*\pFq[6][8]{%
  \begingroup 
  \pFqmuskip=#1mu\relax
  \mathcode`\,=\string"8000
  \begingroup\lccode`\~=`\,
  \lowercase{\endgroup\let~}\pFqcomma
  {}_{#2}F_{#3}{\left[\genfrac..{0pt}{}{#4}{#5};#6\right]}%
  \endgroup
}
\newcommand{\pFqcomma}{\mskip\pFqmuskip}

\section{Introduction}

The notion that degrees of freedom at different length scales decouple from each other
is a cornerstone of modern physics.  In this note, we consider situations where details of the short-distance physics are unknown,
but one is interested in its long-distance effects as parameterized by effective field theory (EFT) coefficients.
In relativistic quantum theories, it is known that ``not anything goes'': if the short-distance physics
part is compatible with causality and unitarity, the low-energy parameters will obey certain inequalities,
discussed notably in \cite{Adams:2006sv}.
In this paper we explore such inequalities in an effort to carve out the allowed space of local and unitary EFTs.

We will consider asymptotically flat space-time, where the S-matrix encodes long-distance or low-energy observables. We will specifically study a subset of EFT parameters, denoted $g_k$, captured by
$2\to2$ scattering. As will be reviewed below, causality and unitarity imply dispersive sum rules:
\be
 g_k = \sum_J \int_{M^2}^\infty ds\ (\cdots)_k\  \prob_J(s)  \label{sum rule intro}
\ee
where the spectral density $\prob_J(s)$ (proportional to the imaginary part of the amplitude)
is related to the probability of a high-energy state with angular momentum $J$ to scatter at energy $\sqrt{s}>M$,
and the kernels $(\cdots)_k$ are given explicitly below and depend on the particular EFT coefficient $g_k$ of interest.
The mass $M$ separates ``light'' and ``heavy'' states and can be interpreted as the EFT cutoff in an appropriate scheme.
We will be agnostic about the high-energy sector: our \emph{only} input will be its compatibility with unitarity and crossing symmetry.
Unitarity, the statement that probabilities should lie between 0 and 1, will simply mean:
\be
0\leq  \prob_J(s) \leq 2\,. \label{unitarity}
\ee
Not all spectral densities that satisfy this inequality are reasonable candidates for the imaginary part of a scattering amplitude,
however. This is because Kramers-Kronig type dispersion relations can reconstruct amplitudes from $\prob_J$ alone, but there is no guarantee
that the outcome satisfies the full crossing symmetry.
Crossing-symmetric $\prob_J(s)$'s are orthogonal to an infinite set of ``null constraints'', which will be a key ingredient of this paper.

For our purposes, classifying causal and unitary EFTs amounts to finding the image,
under the map \eqref{sum rule intro}, of the set of unitary and crossing-symmetric $\prob_J(s)$.

Causality constraints in quantum field theory have been discussed since the inception of the subject.
Many studies were motivated by the phenomenology of the strong force \cite{Martin:1969ina}.
To give just a few examples, dispersion relations and sum rules were used in the analysis
of low-energy pion scattering~\cite{Roy:1971tc,Colangelo:2001df,Caprini:2003ta},
and inequalities satisfied by EFT parameters were obtained using
properties of forward amplitudes in \cite{Pham:1985cr,Ananthanarayan:1994hf}.
This work aims to explore inequalities on EFT parameters systematically.

We focus on the simplest example: a single (non-gravitating) real scalar field.
Since we view the EFT cutoff $M$ as much larger than the mass of the light scattered particles,
we take the latter to be massless.
On grounds of dimensional analysis, one expects the coefficient of a $(k+d)$-dimensional operator
in the low-energy effective Lagrangian to scale like $\sim 1/M^k$, possibly further suppressed by a small coupling, but never larger.
This scaling is clearly realized when one integrates out a massive field.
The main question to be addressed is: \emph{Can dimensional analysis scaling be justified by rigorous numerical bounds?} Can ``accidentally large'' EFT coefficients be ruled out?

We will find that the answer is positive, and we present a general framework to numerically obtain the optimal bounds. Furthermore, we will show that much of the shape of the allowed space, including two kinks,
can be understood from simple analytic scattering amplitudes.

The relation between dimensional analysis scaling and causality
resonates with many previous studies, for example \cite{Camanho:2014apa,Afkhami-Jeddi:2016ntf,Cheung:2016yqr, talks,deRham:2017avq}. Our new observation will be the seemingly universal existence of \emph{two-sided} bounds.

This paper is organized as follows. 
In section \ref{sect:scalar} we review the general principles satisfied by scattering amplitudes, introducing a family of ``$B_k$'' sum rules expressing
EFT coefficients as averages over high-energy probabilities.
In section \ref{sect:semi}, we provide a general numerical optimization strategy to rule-out candidate EFTs by making use of the averaging technology.
In section \ref{sect:numerics}, numerical results are presented along with remarks.
Section \ref{sect:analytic} bridges the numerics with the analytic results.
We conclude in section \ref{sect:conclusion} with a discussion about the potential use cases of the numerical framework presented and the further implications of the numerical results.\\

\noindent {\bf Note added:} When this manuscript was being completed, the works \cite{Bellazzini:2020cot} and then \cite{Tolley:2020gtv} appeared with partial overlap in the results.
The second paper in particular gave a two-sided bound on the $stu$ interaction which agrees with our eq.~\eqref{first lower g3}.
Further comparisons will be interesting.

\section{Preliminaries: Scattering amplitudes and dispersion relations}
\label{sect:scalar}

\subsection{Low energy: effective field theory}

\begin{figure}[t]
    \centering
    \includegraphics[width=0.3\linewidth]{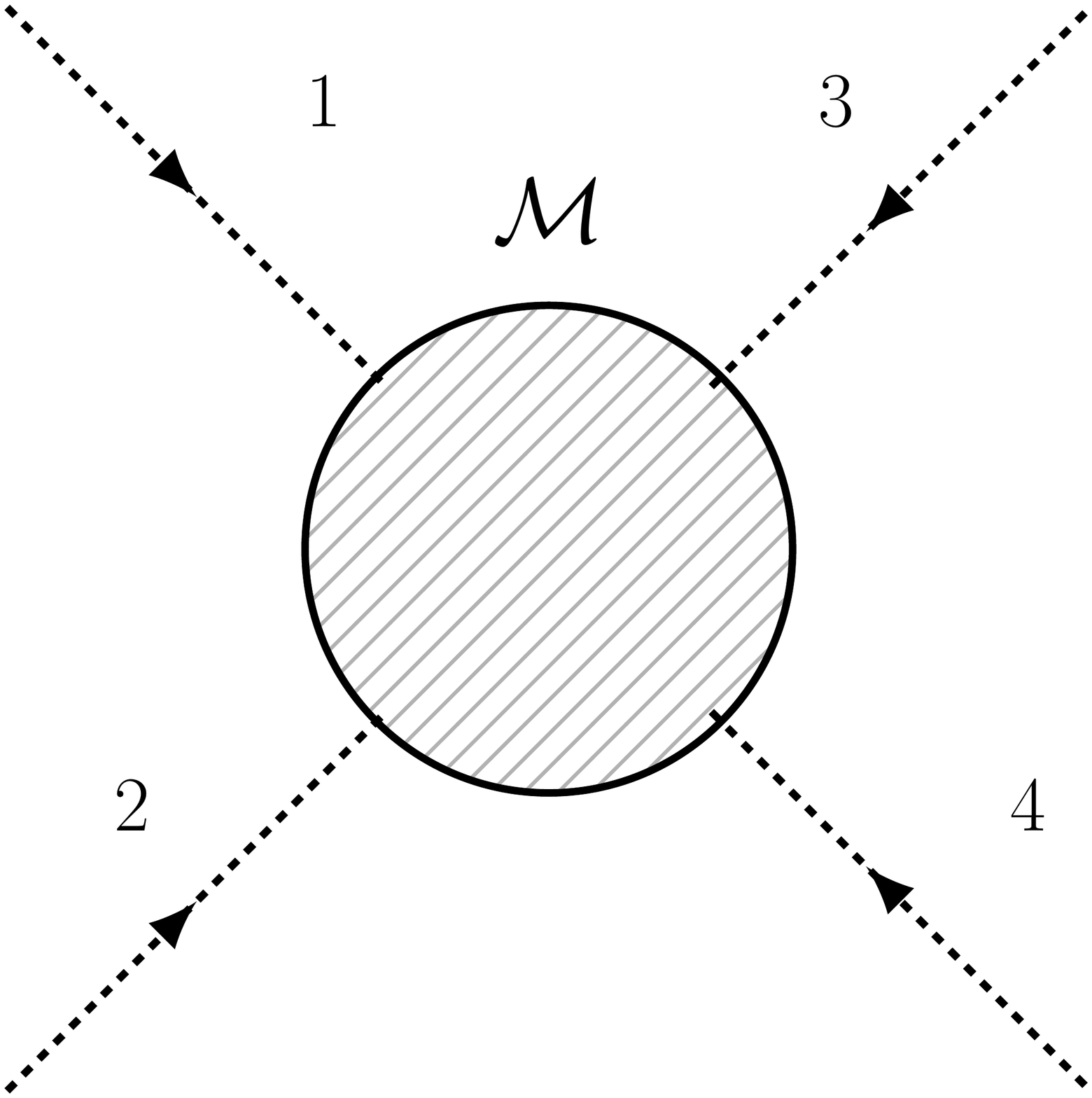}
    \caption{\label{fig:feynman}The $2\to 2$ scattering process studied in this paper.
    For different choices of four-momenta, time can flow either horizontally or vertically (or diagonally).}
    \end{figure}

We consider $2\to 2$ scattering of massless identical real scalars in a Poincar\'e invariant theory
(fig.~\ref{fig:feynman}).
Treating all momenta as incoming, the amplitude is a function of Mandelstam invariants:
\begin{align}
s = -(p_1 + p_2)^2, \quad t = -(p_2+p_3)^2, \quad u = -(p_1 + p_3)^2
\end{align}
which satisfy $s+t+u=0$.
By crossing symmetry, it is invariant under all permutations
(this holds with appropriate $i0$'s in the discontinuity, as further discussed below):
\be
\MM(s,t) = \MM(t,s) = \MM(s,u) = \dots
\ee
Our first step is to parameterize the amplitude at low energies in terms of a specific effective field theory.
Generally, the form of the amplitude depends on the couplings of the theory. It becomes particularly simple if the theory
is weakly coupled and we restrict ourselves to the tree approximation.
We thus use the tree approximation here and
until subsection \ref{ssec:bk_section}  In this case, the amplitude has no low-energy branch-cuts, so the EFT expansion is simply a series in small $s,t,u$:
\begin{align} \begin{split}
\MM_{\text{low}}(s,t) = & -g^2 \left[\frac{1}{s} + \frac{1}{t} + \frac{1}{u}\right] - \lambda\\
& + g_2(s^2 + t^2 + u^2) +g_3 (stu) + g_4(s^2 + t^2 + u^2)^2 + g_5(s^2 + t^2 + u^2)(stu)\\
 & + g_6(s^2 + t^2 + u^2)^3 + g_6^\prime(stu)^2 + g_{7} (s^2 + t^2 + u^2)^2 (stu) + \cdots
\end{split} \label{eqn:eft_amplitude}\end{align}
The first line accounts for $\phi^3$ and $\phi^4$ relevant interactions,
while the remaining terms simply list the most general symmetric polynomials in $s,t,u$,
to account for higher-dimension operators in the EFT.  The subscript denotes the degree in Mandelstam invariants.
Symmetric polynomials are easy to enumerate since their ring
is freely generated by two elements: $s^2+t^2+u^2$ and $stu$ (given that $s+t+u=0$).

A short exercise shows that the preceding amplitude is obtained from the following effective Lagrangian in the tree approximation:
\be\begin{aligned} \label{Llow}
 \mathcal{L}_{\rm low} =& -\frac12(\partial_\mu\phi)^2 - \frac{g}{3!}\phi^3 -\frac{\lambda}{4!}\phi^4
\\ & +\frac{g_2}{2} \big[(\partial_\mu\phi)^2\big]^2+\frac{g_3}{3} (\partial_\mu\partial_\nu\phi)^2(\partial_\sigma\phi)^2
+4g_4\big[(\partial_\mu\partial_\nu\phi)^2\big]^2
+\cdots
\end{aligned}\ee
As is well known,  Lagrangian densities are not unique: they are defined modulo integration-by-parts
and field redefinitions.
One can cast any effective Lagrangian for a real scalar field into the form \eqref{Llow} by using field redefinitions to
eliminate, order by order in the derivative expansion,
corrections to the kinetic and cubic terms as well as appearances of $\partial^2\phi$.
See for example \cite{Brivio:2017vri} for a discussion in the Standard Model context.
The amplitude \eqref{eqn:eft_amplitude} is a physical observable unaffected by such ambiguities,
which is why we choose to parameterize the coefficients in terms of it.

Our goal is to constrain the EFT parameters $g_k$ assuming existence of an high-energy completion which is
causal and unitary, but not necessarily weakly coupled.
Low-energy interactions involving five or more powers of $\phi$ will not be constrained by our methods,
since they are not detected by (tree-level) $2\to2$ scattering.
When low-energy loop corrections are included, the detailed form of eq.~\eqref{eqn:eft_amplitude} will be
modified, but we do not expect the number of independent EFT parameters that we can constrain
to increase. A precise definition of the $g_k$'s that remains valid in the presence of low-energy loop corrections
is given in eq.~\eqref{Bks general} below.

\subsection{High energy: partial wave decomposition}
\label{sect:gegenbauer}

At high energies, we will be agnostic about the form of the amplitude except for the assumption that it is causal and unitary.  We follow the general framework of S-matrix theory, as reviewed for example in \cite{Martin:1969ina}.
Let us begin with unitarity of the S-matrix, which is formally that: $S^\dagger S=1$, where $S=1+i\mathcal{M}$. More precisely, one picks a physical region, say where $s>0$ is interpreted as center-of-mass energy squared, and
$-s<t<0$ gives the momentum transfer (squared); the scattering angle is
\be
 \cos\theta = 1+ \frac{2t}{s}\,.
\ee 
The scattering operator is a convolution with respect to angles, which is diagonalized by going to a basis of angular momentum partial waves.
The unitarity condition is thus simplest to state in this basis (our conventions follow \cite{Correia:2020xtr}):
\be \MM(s,t) = 
\sum_{J \text { even }}
 n_{J}^{(d)} f_J(s)  \legP_J\left(1+\tfrac{2 t}{s}\right),
\quad n_{J}^{(d)}=\frac{(4 \pi)^{\frac{d}{2}}(d+2 J-3) \Gamma(d+J-3)}{\pi \Gamma\left(\frac{d-2}{2}\right) \Gamma(J+1)}\,,
\label{eq:gegenbauer_expansion}
\ee
where $d$ is the space-time dimensions and $\legP_J(x)$ are the $d$-dimensional version of Legendre polynomials (which appear in the $d=4$ case):
\be
\legP_{J}(x) \equiv{ }_{2} F_{1}\left(-J, J+d-3, \frac{d-2}{2}, \frac{1-x}{2}\right)\,. \label{gegenbauer}
\ee
Details of the specific scattering process are encoded in the coefficients $f_J(s)$.
In this normalization convention, unitarity of the elastic amplitude of identical real particles is \cite{Correia:2020xtr}:
\be
 \big|S_J(s)\big|\leq 1, \qquad S_J(s) \equiv1+ i s^\frac{d-4}{2} f_J(s)\,.
\ee
The elastic amplitude $S_J$ can have absolute value less than unity due to inelastic processes.
We will only need the imaginary part of the high-energy amplitude.
Defining the spectral density $\prob_J(s)=s^\frac{d-4}{2} {\rm Im\ }f_J(s)$, it can thus be written as
\be
{\rm Im}\ \MM(s,t) = s^{\frac{4-d}{2}} \sum_{J \text { even }} n_{J}^{(d)} \  \prob_J(s)  \legP_J\left(1+\tfrac{2 t}{s}\right)
\label{Im partial waves}
\ee
where the unitarity constraint is
\be
 0\leq \prob_J(s)\leq 2 \qquad \forall s>0, \forall J\ {\rm even}\,. \label{eqn:positivity_gegenbauer}
\ee
The normalization is such that $\prob_J=1$ for complete absorption ($S_J=0$), and $\prob_J=2$ for an elastic phase shift $\pi$.
For the most part (except for subsection \ref{ssec:g2}) we will only use the first inequality: $ 0\leq \prob_J(s)$.

\subsection{Dispersion relations}

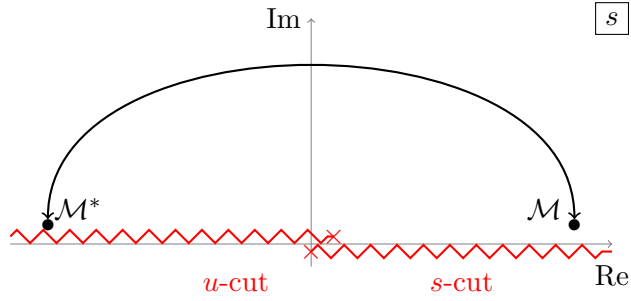
\begin{figure}[t]
\centering
\scalebox{1}{
\begin{tikzpicture}
  [
    decoration={%
      markings,
      mark=at position 0.5 with {\arrow[line width=1pt]{>}},
    }
  ]
  \draw [help lines,->] (-4,0) -- (4,0) coordinate (xaxis);
  \draw [help lines,->] (0,-0.3) -- (0,3) coordinate (yaxis);
  \node [below=0.15cm] at (xaxis) {Re};
  \node [left] at (yaxis) {Im};

  \node[red] at (0,-0.1) {$\times$};
  \node[red] at (0.3,+0.1) {$\times$};
  
  \node[black] at (-3.5,0.25) {\textbullet};
  \node[black] at (3.5,0.25) {\textbullet};

  \node[below=0.25cm,red] at (-1,0) {$u$-cut};
  \node[below=0.25cm,red] at (2,0) {$s$-cut};
  
  \node[above=0.2cm] at (-3.1,0) {$\mathcal{M}^*$};
  \node[above=0.2cm] at (3.1,0) {$\mathcal{M}$};
  
  \draw[<->,bend angle=90,thick] (-3.5,0.33) to[bend left] (3.5,0.33);
  
  \draw[line width=0.8pt, red] (-4,0.1) [decorate, decoration=zigzag] --(0.3,0.1);
  \draw[line width=0.8pt, red] (0,-0.1) [decorate, decoration=zigzag] --(4,-0.1);
  
  \node[draw] at (4,3) {$s$};
  
\end{tikzpicture}
}
   \caption{\label{fig:analyticity}
Analyticity in the upper-half-plane relates the $s$-channel amplitude and the complex conjugate (anti-time-ordered)
$u$-channel amplitude. Note that the $s$- and $u$-channel cuts overlap in the physical region where $t<0$, which is not a problem
since the crossing path avoids small $|s|$.}
\end{figure}

The other key ingredient from S-matrix theory is the connection between low and high energies,
which stems from analyticity.
More precisely, we will use the following two properties of the amplitude:
\begin{enumerate}
\item For fixed $t<0$ and $|s|$ sufficiently large, $\MM(s,t)$ is analytic in $s$ away from the real axis.
\item For fixed $t<0$, $\lim_{|s|\to\infty} \left|\frac{\MM(s,t)}{s^2}\right|=0$ along any line of constant phase.
\end{enumerate}
Physically, these conditions combine causality and unitarity.
For an elementary explanation of their respective significance, we refer to the signal propagation model in appendix D of \cite{Camanho:2014apa},
where it is explained that propagation of a signal through a black box is causal if and only if the corresponding transfer function
$S(\omega)$ is analytic in the upper-half frequency plane (with sub-exponential growth),
and that $|S(\omega)|^2\leq 1$ \emph{throughout the upper-half-plane} if the box furthermore preserves the squared-norm of signals.
These are general facts about Fourier transforms.
The original physical derivation of crossing symmetry \cite{GellMann:1954db} applies these facts to the expectation value of a retarded commutator
in one-particle state. Schematically, one considers 
\be
 \langle p_4|\ [\phi(x_3),\phi(x_2)]\theta(x_3^0-x_2^0)\ |p_1\rangle,
\ee
which vanishes outside the forward light-cone rendering its Fourier transform analytic in the upper-half $s$-plane
(intuitively, one uses that $s$ is linear in right-moving light-cone momentum), at least for large enough $|s|$.
Its boundary values on the real axis unite the $s$-channel amplitude
and the complex conjugate of the $u$-channel amplitude, see fig.~\ref{fig:analyticity}.

This is the traditional understanding of crossing symmetry within the axiomatic theory \cite{Bros:1965kbd}.
The boundedness property, in particular including the extra factor of $1/s^2$ compared with the signal model, will be critical for us.
We believe it can be justified physically by directly analyzing the transverse Fourier transform \cite{SCH:bootstrap2020}.
As far as we understand, properties 1-2 are theorems in axiomatic quantum field theory, for example in the context of pion scattering \cite{Jin:1964zza,Martin:1965jj}; we take them as axioms embodying causality and unitarity.

The two properties assumed above amount to the existence of twice-subtracted dispersion relations.
Let us derive such dispersion relations explicitly.
The starting point is that an integral over a large circle vanishes:
\be
  \oint_{\infty} \frac{ds'}{2\pi i(s'-s)} \frac{\MM(s',t)}{(s'-s_1)(s'-s_2)} = 0 \label{UV zero}
\ee
where $s_1$ and $s_2$ are arbitrary subtraction points. For large enough $s'$, the integrand behaves like 
$\sim\MM(s',t)/s'^3$, and so the integral vanishes thanks to property 2.
Typically, one would formally treat all of $s$, $s_1$ and $s_2$ as non-real and deform the contour toward the real axis.
Summing the three explicit poles and cuts then relates $\MM(s,t)$ to its value at two subtraction points plus
an integral over the discontinuity of $\MM$ across the real axis \cite{Eden:1966dnq}.

For our purposes it is convenient to instead treat the subtraction poles as part of the real-axis cuts.
We choose $s_1=0$ and $s_2=-t$ to maintain the symmetry
between the $s$ and $u$ channels without introducing any new energy scale into the problem.
The identity \eqref{UV zero} then relates the residue at $s'=s$ with a discontinuity:
\be
 \frac{\MM(s,t)}{s(s+t)} = \int_{-\infty}^{\infty} \frac{ds'}{\pi(s'-s)} \ {\rm Im}\left[\frac{\MM(s',t)}{s'(s'+t)}\right] \qquad (t<0, s\notin \mathbb{R}),
 \label{dispersion relation}
\ee
where we have written the discontinuity as an imaginary since the amplitude on the ``wrong side of the cut'' is its complex conjugate;
technically ${\rm Im}\ f(s) \equiv \tfrac{1}{2i}\big[f(s+i0)-f(s-i0)\big]$.
We call eq.~\eqref{dispersion relation} a
twice-subtracted dispersion relation because of the two powers of $s'$ added to the denominator.

\begin{figure}[t]
	\centering
    \begin{subfigure}[t]{0.45\textwidth}
     \centering
	\raisebox{-2mm}{\scalebox{0.83}{
  \begin{tikzpicture}
  [
    decoration={%
      markings,
      mark=at position 0.5 with {\arrow[line width=1pt]{>}},
    }
  ]
  \draw [help lines,->] (-4,0) -- (4,0) coordinate (xaxis);
  \draw [help lines,->] (0,-3) -- (0,3) coordinate (yaxis);
  \node [right] at (xaxis) {Re};
  \node [left] at (yaxis) {Im};

  \node[red] at (1,0) {$\times$};
  \node[red] at (0,0) {$\times$};
  \node[red] at (2,0) {$\times$};
  \node[red] at (-1,0) {$\times$};

  \draw[line width=0.8pt, red] (-4,0) [decorate, decoration=zigzag] --(-1,0);
  \draw[line width=0.8pt, red] (2,0) [decorate, decoration=zigzag] --(4,0);

  \node[below=0.25cm,red] at (1,0) {$-t$};
  \node[above=0.25cm,red] at (-1.5,0) {$-M^2 - t$};
  \node[above=0.25cm,red] at (2,0) {$M^2$};
  \node[below=0.25cm,black] at (-3,3) {$|s|\rightarrow \infty$};
  
 \path [draw, line width=0.8pt, postaction=decorate,blue,bend angle=90] (-4,0.2) to[bend left] (4,0.2);
 \path [draw, line width=0.8pt, postaction=decorate,blue,bend angle=90] (4,-0.2) to[bend left] (-4,-0.2);

 \node[draw] at (4,3) {$s$};

\end{tikzpicture}
}}
    \end{subfigure}\hspace{0mm}
\raisebox{25mm}{$\ \ =$}
    \begin{subfigure}[t]{0.45\textwidth}
\centering
\scalebox{0.83}{
\begin{tikzpicture}
  [
    decoration={%
      markings,
      mark=at position 0.5 with {\arrow[line width=1pt]{>}},
    }
  ]
  \draw [help lines,->] (-4,0) -- (4,0) coordinate (xaxis);
  \draw [help lines,->] (0,-3) -- (0,3) coordinate (yaxis);
  \node [right] at (xaxis) {Re};
  \node [left] at (yaxis) {Im};

  \node[red] at (1,0) {$\times$};
  \node[red] at (0,0) {$\times$};
  \node[red] at (2,0) {$\times$};
  \node[red] at (-1,0) {$\times$};

  
  \draw[line width=0.8pt, red] (-4,0) [decorate, decoration=zigzag] --(-1,0);
  \draw[line width=0.8pt, red] (2,0) [decorate, decoration=zigzag] --(4,0);
  
  \path [draw, line width=0.8pt, postaction=decorate,blue] (-0.2,0) arc (180:0:.2);
  \path [draw, line width=0.8pt, postaction=decorate,blue] (.2,0) arc (0:-180:.2);
  
  \path [draw, line width=0.8pt, postaction=decorate,blue] (0.8,0) arc (180:0:.2);
  \path [draw, line width=0.8pt, postaction=decorate,blue] (1.2,0) arc (0:-180:.2);
  
  \node[below=0.25cm,red] at (1,0) {$-t$};
  \node[above=0.25cm,red] at (-1.5,0) {$-M^2 - t$};
  \node[above=0.25cm,red] at (2,0) {$M^2$};
  
  \path [draw, line width=0.8pt, postaction=decorate,blue] (-4,0.2) -- (-1,0.2);
  \path [draw, line width=0.8pt, postaction=decorate,blue] (-1,0.2) arc (90:-90:0.2) -- (-4,-0.2);
  
  \path [draw, line width=0.8pt, postaction=decorate,blue] (4,-0.2) -- (2,-0.2);
  \path [draw, line width=0.8pt, postaction=decorate,blue] (2,-0.2) arc (270:90:0.2) -- (4,0.2 );
  
  \node[draw] at (4,3) {$s$};
\end{tikzpicture}
}
   \end{subfigure} 
   \caption{\label{fig:contour_deformation}
Contour deformation which gives the sum rule \eqref{disp low high} when low-energy loops are neglected:
the integral over arcs at infinity vanishes, thus relating low-energy data and heavy cuts.}
\end{figure}
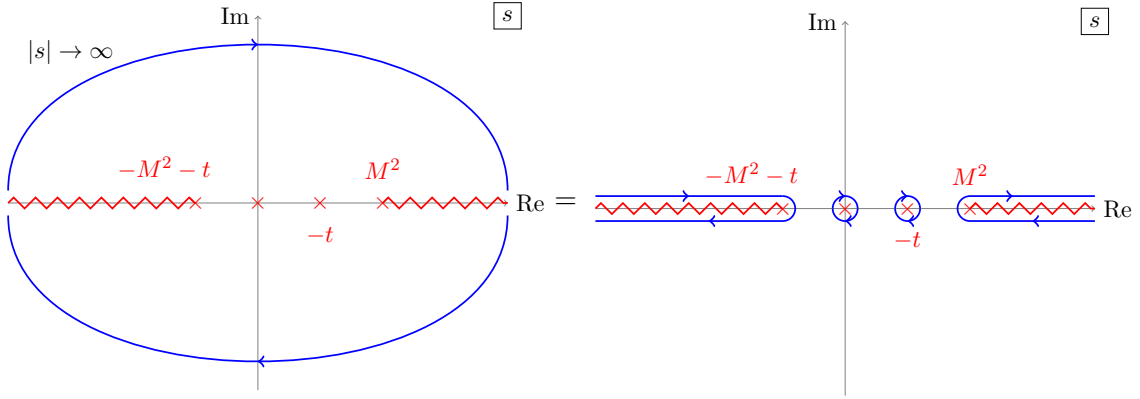

Let us see how this works in the simplest situation considered in eq.~\eqref{eqn:eft_amplitude},
where low-energy loops are neglected. Then branch cuts can only start at the UV cutoff $M^2$.
The right-hand-side of eq.~\eqref{dispersion relation} then contains low-energy poles at $s=0$ and $s=-t$
(due both to the denominator in eq.~\eqref{dispersion relation} and poles in the amplitude),
and high-energy cuts at $s>M^2$ and $s<-M^2-t$.
Separating low and high energies gives a relation:
\be\begin{aligned}
& \frac{\MM_{\rm low}(s,t)}{s(s+t)} + \Res\limits_{s'=0,-t} \left[\frac{1}{s'-s}\frac{\MM_{\rm low}(s',t)}{s'(s'+t)}\right]
\\ & \qquad = \int_{M^2}^\infty \frac{ds'}{\pi} \left(\frac{1}{s'-s} + \frac{1}{s'+s+t}\right) \ {\rm Im}\left[\frac{\MM_{\rm high}(s',t)}{s'(s'+t)}\right]\, .
\end{aligned}\label{disp low high}\ee
We used $s\leftrightarrow u$ symmetry to combine the left and right cuts.
This relation is supposed to converge for any $s,t$ with $u<0$.
Interestingly, plugging in the EFT expansion eq.~\eqref{eqn:eft_amplitude} for $\MM_{\rm low}$,
one finds that both the spin-0 exchange diagram $g^2$ and spin-0 contact interaction $\lambda$ cancel out,
and what remains is pole-free (this could have been anticipated from the fact that the three residues on the left combine into a single contour over a large circle).
On the right-hand-side we insert the partial wave expansion \eqref{Im partial waves}.
It is useful to define heavy averages:
\be \label{avg def}
\avg{F(m^2,J)} \equiv \sum_{J \text { even }} n_{J}^{(d)} \int_{M^2}^{\infty} \frac{dm^2}{\pi} \frac{m^{4-d}}{m^2} \prob_J(m^2) \left[F(m^2, J)\right].
\ee
Eq.~\eqref{disp low high} becomes, for $t<0$:
\be
2 g_2 - t g_3 + 4\big(2t^2+s(s+t)\big) g_4 +\ldots\quad = \avg{ \frac{2m^2+t}{(m^2-s)(m^2+s+t)} \frac{\legP_J(1+\frac{2t}{m^2})}{m^2(m^2+t)}}\,.
\label{dispersion}
\ee
The averaging symbol denotes a (non-normalized) positive sum over heavy states with mass $m>M$.
All the results in this paper follow from Taylor-expanding both sides in $s$ and $t$ and using positivity of the measure
$\avg{\cdots}$. It will be useful (though non-essential) to re-organize a bit.

\subsection{The $B_k(t)$ family of sum rules}

\label{ssec:bk_section}

It is easy to see that the Taylor expansion of both sides of eq.~\eqref{dispersion} maintains the symmetry under $s\to -s-t$,
and therefore only even powers of $s$ carry information.
More precisely, for each even integer $k$ the coefficient of $[s(s+t)]^{k/2-1}$ gives a one-parameter family of sum rules parameterized by $t$,
which we call $B_{k}(t)$.  It can be computed by taking the $s\to 0$ limit in eq.~\eqref{UV zero}:
\be \label{Bk def}
 B_k(t)\equiv \oint_{\infty} \frac{ds}{2\pi i} \frac1s\ \frac{\MM(s,t)}{\big[s(s+t)\big]^{k/2}} =0 \qquad\qquad (t<0,\ k=2,4,\ldots)\,.
\ee
This is similar to moment sum rules $\oint_{\infty} \frac{ds}{s^{k+1}}\MM(s,t)$ which have been used since times immemorial.
Here we have simply re-organized using the $s\leftrightarrow u$ symmetry of our problem to eliminate odd moments.\footnote{
The identity: $\oint_\infty \frac{ds}{2\pi i} \frac{1}{s}\big[s(s+t)\big]^{(k-k')/2}=\delta_{k,k'}$
shows that eq.~\eqref{Bk def} indeed extracts the coefficient of $[s(s+t)]^{k/2}$ in $\MM_{\rm low}$.
}
The subscript indicates that $B_k$ enjoys the high-energy convergence of a $k$-subtracted dispersion relation.

A closely related basis of sum rules was introduced recently for conformal field theory correlators \cite{Caron-Huot:2020adz} (see also \cite{Penedones:2019tng}).
For holographic theories, the Mellin-space form of the sum rule, called $\widehat{B}_{k,t}$ (see eq.~(4.54) and section 4.8 there),
precisely reduces in the flat space limit to our current $B_{k}(t)$.
In this context, convergence for $k\geq 2$ is a consequence of the known boundedness of conformal correlators in the Regge limit.

\begin{figure}[t]
\centering
\scalebox{1}{
\begin{tikzpicture}
  [
    decoration={%
      markings,
      mark=at position 0.5 with {\arrow[line width=1pt]{>}},
    }
  ]
  \draw [help lines,->] (-4,0) -- (4,0) coordinate (xaxis);
  \draw [help lines,->] (0,-2) -- (0,2) coordinate (yaxis);
  \node [below=0.25cm] at (xaxis) {Re};
  \node [left] at (yaxis) {Im};
  
   \node[red] at (0.2,0) {$\times$};
  \node[red] at (0,0) {$\times$};
  \node[red] at (1,0) {$\times$};
  \node[red] at (-1,0) {$\times$};

  \draw[line width=0.8pt, red] (-4,0) [decorate, decoration=zigzag] --(-1,0);
  \draw[line width=0.8pt, red] (1,0) [decorate, decoration=zigzag] --(4,0);
  
  
  
  \node[below=0.25cm,red] at (0.2,0) {$-t$};
  \node[below=0.25cm,red] at (-1.75,0) {$-M^2 - t$};
  \node[below=0.25cm,red] at (1.5,0) {$M^2$};
  
  \path [draw, line width=0.8pt, postaction=decorate,blue] (-4,0.2) -- (-1,.2); \path [draw, line width=0.8pt, postaction=decorate,blue] (-1,0.2) arc (180:0:1);
  \path [draw, line width=0.8pt, postaction=decorate,blue] (1,.2) -- (4,0.2);
  
  \path [draw, line width=0.8pt, postaction=decorate,blue] (4,-0.2) -- (1,-.2); \path [draw, line width=0.8pt, postaction=decorate,blue] (1,-0.2) arc (0:-180:1);
  \path [draw, line width=0.8pt, postaction=decorate,blue] (-1,-.2) -- (-4,-0.2);

  \node[draw] at (4,2) {$s$};
  
\end{tikzpicture}
}
\caption{\label{fig: good Bk}
Integration contour to be used when low-energy loops are included; the integral vanishes as it is equivalent to arcs at infinity.
This relates high-energy cuts at $s>M^2$ and $u>M^2$ with EFT-computable data near the EFT cutoff $|s|\sim M^2$.}
\end{figure}
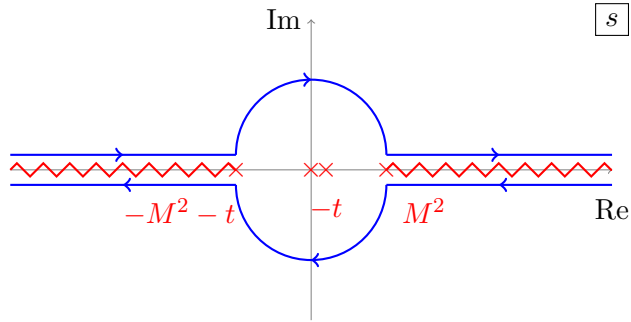

For massless scattering, the low-energy $s$- and $u$-channel cuts of $\MM(s,t)$ generally overlap as shown in fig.~\ref{fig:analyticity}.
It is important that eq.~\eqref{Bk def} can be computed \emph{without} going between the cuts.
We simply deform the contour to pick heavy branch cuts at $s>M^2$ and $u>M^2$, and keep the
rest as large arcs with $|s|\sim M^2$, the EFT cutoff, see fig.~\ref{fig: good Bk}.
This gives a relation between physics at the scale $M$ and that at higher-energies:
\be
\label{Bks general}
B_k:\ \oint_{|s|\approx M^2} \frac{ds}{2\pi i} \frac1s \frac{\MM(s,t)}{\big[s(s+t)\big]^{k/2}}
	= \avg{ \frac{2m^2 + t}{m^2+t} \frac{\legP_J^{(d)}\left(1+\tfrac{2 t}{m^2}\right)}{\big[m^2(m^2+t)\big]^{k/2}}}\qquad (t<0,\ k=2,4,\ldots)\,.
\ee
This equation is valid even when EFT loops are included.
The idea is to choose the EFT cutoff $M$ such that loop corrections in the low-energy EFT are under control
over the arcs with $|s|\approx M^2$.  Eq.~\eqref{Bks general} thus equates an EFT-computable left-hand-side, with a high-energy average that enjoys positivity properties.

The specific relation between the left-hand-side and EFT coefficients will depend on EFT interactions.
For concreteness let us thus focus again on the case where EFT loops are neglected.
The left-hand-side is then just the sum of residues at $s=0$ and $s=-t$; 
By symmetry, we can replace $\frac1s$ by $\left(\frac{1}{s}-\frac{1}{s+t}\right)$ and include a single pole,
and the $B_k$ sum rules becomes:
\be
\label{Bks}
\boxed{B_k:\ \Res\limits_{s=0} \left[\frac{2s + t}{s(s+t)}\frac{\MM_\text{low}(s,t)}{\big[s(s+t)\big]^{k/2}}\right]
	= \avg{ \frac{2m^2 + t}{m^2+t} \frac{\legP_J^{(d)}\left(1+\tfrac{2 t}{m^2}\right)}{\big[m^2(m^2+t)\big]^{k/2}}}\qquad (t<0,\ k=2,4,\ldots)\,.
}
\ee
This simplification of eq.~\eqref{Bks general} is only valid when neglecting EFT loops.

Let us record the first few two instances explicitly:
\begin{align}
       B_2: &\quad 2 g_{2}-g_{3} t+8 g_{4} t^2+\ldots& = \left\langle\frac{\left(2 m^{2}+t\right) \legP_J\left(1+\frac{2 t}{m^2}\right)}{m^{2}\left(m^{2}+t\right)^{2}}\right\rangle\,,\\
       B_4: &\quad 4g_4 + \ldots & = \left\langle\frac{\left(2 m^{2}+t\right) \legP_J\left(1+\frac{2 t}{m^{2}}\right)}{m^{4}\left(m^2+t\right)^{3}}\right\rangle\,.
\end{align}
The left-hand side has a regular series in $t$, and the right-hand side involves Gegenbauers $\legP_J(1+\tfrac{2t}{m^2})$,
which can be straightforwardly expanded at small $t\ll M^2$ using eq.~\eqref{gegenbauer}\,.
Recall that averages are taken over heavy states with $m\geq M$.
Matching both sides order by order in $t$ generates a linear system in $g_n$'s:
\def\JJ{\mathcal{J}}
\be\begin{aligned}
\label{eq:list_of_gs}
	g_2 &=  \avg{\frac{1}{m^4}}\,, \qquad
	g_3 = \avg{\frac{3-\frac{4}{d-2}\JJ^2}{m^6}}\,, \qquad
	g_4 = \avg{\frac{1}{2m^8}}\,,\\
	g_4 &= \avg{ \frac{1 + \frac{4-5d}{2d(d-2)}\JJ^2 + \frac{1}{d(d-2)}\JJ^4}{2m^8} }\,.
\end{aligned}\ee
We introduced the spin Casimir $\JJ^2 = J(J+d-3)$ for convenience.  Note that we truncated $\mathcal{M}_{\text{low}}$ to order $g_4$, but it is possible to work to higher orders and generate linear relations on couplings such as $g_5$ and so on. 

The averaging notation immediately shows that $g_2,g_4>0$ since they are high-energy
averages of positive quantities $\frac{1}{m^4}$ and $\frac{1}{2m^8}$, respectively. Furthermore, the inequalities
$g_3\leq \frac{3g_2}{M^2}$ and $g_4\leq \frac{g_2}{2M^4}$ also follow readily since $m\geq M$ inside the average.
In contrast, the sign of $g_3$ is not immediate due to the presence of spinning particles -- the magnitude of $\mathcal{J}^2$ requires a deeper investigation.  
This difficulty was noted in attempted proofs of the six-dimensional $a$-theorem \cite{Elvang:2012st}.


\begin{figure}[t]
    \centering
    \includegraphics[width=0.6\textwidth]{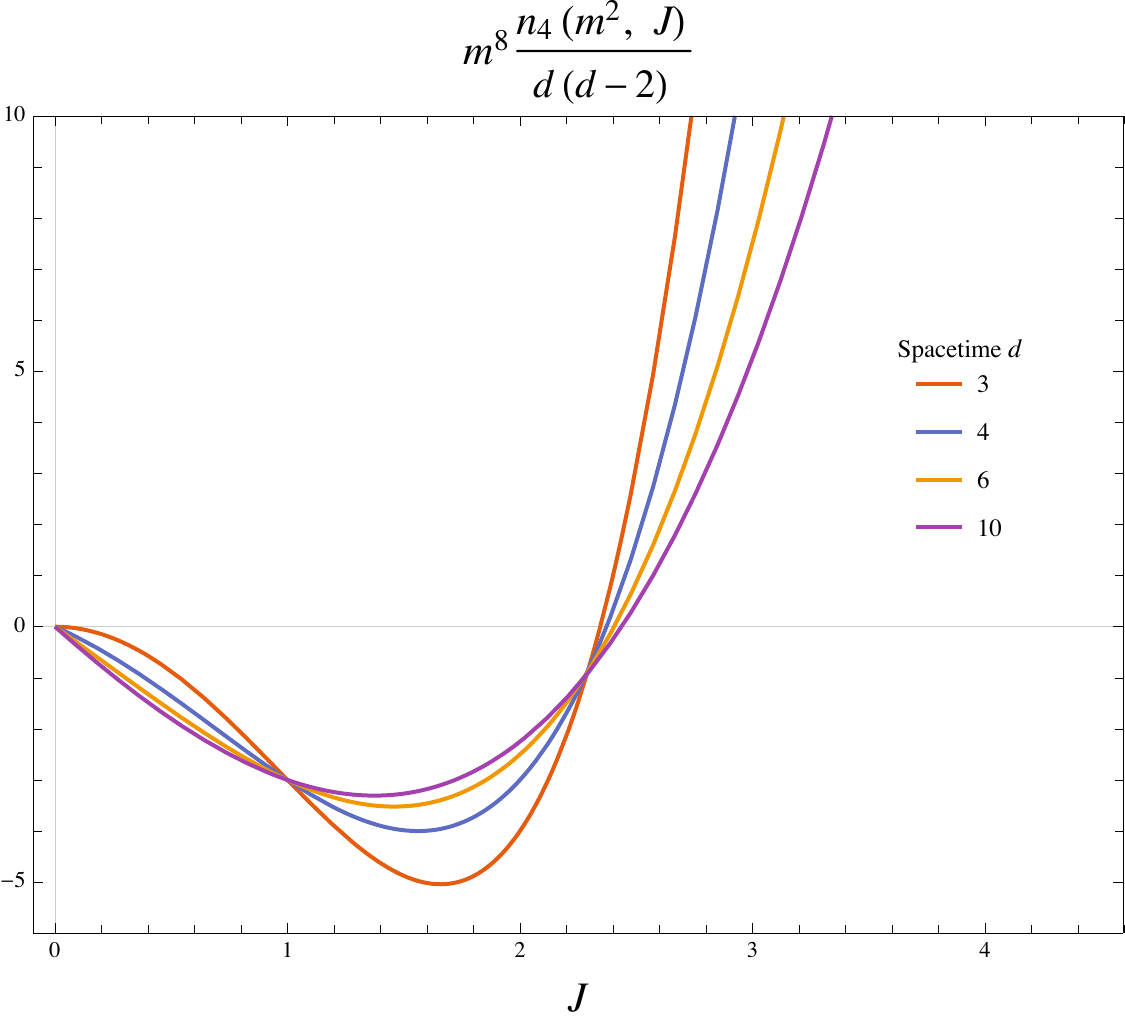}
    \caption{The null constraint $m^8 n_4(m^2,J)$, which is a function of only $J$.
    It vanishes at $J=0$, is negative at $J=2$, but positive at $J=4,6,\dots$, and thus balances spin-two against higher-spin states.
    The sign change in various space-time dimensions (at $J_{\text{critical}} = \frac12(3-d + \sqrt{d (d+4)+1})$)
    is always situated between $2<J<3$. }
    \label{fig:null_plot}
    \end{figure}

The key to calculating a lower bound for $g_3$ will be the existence of two distinct averages that output $g_4$.
Equating them yields the first example of what turns out to be an infinite set of \emph{null constraints}:
\be
     0 = \avg{n_4(m^2,J)}, \qquad    n_4(m^2,J) \equiv \frac{\JJ^2 \big(2\JJ^2 -(5d-4)\big)}{m^8}\,.
\label{eqn:null1}
\ee
This is a constraint on the probabilities $\prob_J(s)$ which define the average $\avg{\cdot}$.  The subscript indicates the degree in $1/m^2$.
Physically this stems from crossing symmetry -- since there is a unique symmetric polynomial at degree 4,
the coefficients of $s^2t^2$ and $s^4$ must be related. There are no lower-degree examples of this phenomenon:
monomials with fewer than two powers of $s$ are killed by any double-subtracted sum rule,
and odd powers of $s$ are information-free since fixed-$t$ dispersion relations preserve the $s\leftrightarrow -s-t$ symmetry of our problem.

Null constraints such as eq.~\eqref{eqn:null1} will be central to this work.
They balance spin-two states against higher spin states:
as visible from fig.~\ref{fig:null_plot}, the average vanishes for spin 0, is negative for spin 2, and positive for all other spins.
This implies that, as soon as one particle of spin 2 is present, higher-spin particles must also be present, with predictable properties.
(Spin two particles are singled out by the physical assumption that double-subtracted sum rule converges.)


\section{Optimization framework}\label{sect:semi}

\label{sect:optimization}

The $B_k$ sum rules just introduced, coupled with positivity of high-energy
averages $\avg{\cdot}$ provide a complete apparatus to establish potent self-consistency conditions on EFT coefficients $g_k$'s (defined in eq.~\eqref{eqn:eft_amplitude}).  We recall our physical assumptions:
\begin{itemize}
\item Double-subtracted dispersion relations converge
\item The low-energy amplitude is crossing symmetric
\item The high-energy spectral density is positive
\end{itemize}
Since we are considering averages over heavy states (with $m>M$), the coefficients (except in subsection \ref{ssec:g2}) are naturally normalized by $g_2$ and the EFT cutoff $M$. We will therefore be bounding dimensionless ratios:
\begin{align}
	\tilde{g}_3 = g_3 \frac{M^2}{g_2}, \qquad \tilde{g}_4 = g_4 \frac{M^4}{g_2}, \qquad \tilde{g}_5 = g_5 \frac{M^6}{g_2},  \quad \dots
\label{eqn:normalized_couplings}
\end{align}
Optimal bounds on these $\tilde{g}_k$'s will be found by formulating a \emph{dual problem}, in which we combine the desired averages (such as \ref{Bks general}) with null constraints (such as eq.~\eqref{eqn:null1})
to obtain sign-definite sum rules.  We first describe a simple example analytically, then describe a systematic implementation
as a semi-definite problem amenable to publicly available software like \texttt{SDPB} \cite{Simmons-Duffin:2015qma}.

\subsection{Warm-up problem with three sum rules}

As a warm-up, let us ask whether it is possible to lower-bound the $\tilde{g}_3$ coefficient using the $B_2, B_4$ sum rules previously calculated.
We consider the corresponding system of three equations from \eqref{eq:list_of_gs}
(including the null constraint obtained via $g_4$ data):
\be
       g_2 = \avg{\frac{1}{m^4}},\qquad
       g_3 = \avg{\frac{3-\frac{4}{d-2}\mathcal{J}^2}{m^6}},\qquad
       0    =  \avg{\frac{\mathcal{J}^2 (2 \mathcal{J}^2-5d+4)}{m^8}}\,. \label{warm-up}
\ee
With these definitions, let us examine a similar, but simpler set of relations:
\be
       h_2 = \avg{\frac{1}{m^4}},\qquad
       h_3 = \avg{\frac{a-\mathcal{J}^2}{m^6}},\qquad
       0    =  \avg{\frac{\JJ^4-b\JJ^2}{m^8}}\equiv \avg{n(m^2,J)}\,. \label{warm-up 2}
\ee
These relations take on the same form as original identities when $a =\frac{3(d-2)}{4}$, $b = \frac{5d-4}{2}$ and the coupling is rescaled to $g_3=\frac{4}{d-2}h_3$.  Consequently, our warm-up problem is to lower-bound $h_3$.

What makes a finite lower bound plausible is that the null constraint (the third equation) should somehow prevent large spins from contributing too much.  This is an important point: the allowed range for $\tilde{g}_3$ is restricted
by \emph{higher derivative} crossing equations!

We now calculate a lower bound in two ways.  The first -- and simplest -- method is to use the Cauchy-Schwarz inequality with the null constraint:
\be
 \avg{ \frac{\JJ^2}{m^6}}^2 \leq  \avg{ \frac{1}{m^4}}\avg{\frac{\JJ^4}{m^8}} = b \avg{ \frac{1}{m^4}}\avg{\frac{\JJ^2}{m^8}}\,.
\ee
Then, using the fact that $m>M$ and $\JJ^2\geq 0$ inside the average yields
$\avg{\frac{\JJ^2}{m^8}} \leq \frac{1}{M^2}\avg{\frac{\JJ^2}{m^6}}$, and by dividing both sides by that average, we obtain an upper-bound on $\avg{\frac{\JJ^2}{m^6}}$ as desired:
\be
 \avg{\frac{\JJ^2}{m^6}} \leq \frac{b}{M^2} \avg{\frac{1}{m^4}} \quad\Rightarrow\quad h_3\geq  -\frac{b}{M^2} h_2\,.
 \label{warm up h3 bound}
\ee
This has a simple physical interpretation: if we define an impact parameter $\tilde{b}=\frac{2J}{m}$, then we have effectively shown that
heavy states can't contribute at impact parameters much larger than $\sim 1/M$.
In terms of the original problem \eqref{warm-up}, we have shown that
\be
 -\frac{2(5d-4)}{d-2} \leq \frac{g_3 M^2}{g_2} \leq 3\,. \label{first lower g3}
\ee
This shows the existence of two-sided bounds for generic couplings.  This is an important qualitative result, to our knowledge
originally emphasized in \cite{talks}:
ratios of EFT couplings, in units of the cutoff scale $M$, \emph{must}
be $\cO(1)$ numbers. Numerically, however, the Cauchy-Schwarz method does not yield the optimal lower bound.

In contrast, the second -- and more powerful -- method is re-interpret the above task as a semi-definite problem, in order to systematically search for optimal bounds.
Denote $h_i(m^2,J)$ the function whose average gives $h_i$.
The idea is to construct positive-definite combinations of the three averages in eq.~\eqref{warm-up 2}:
\be\begin{aligned}
 F(m^2,J) &\equiv h_3(m^2,J) + \alpha h_2(m^2,J) + \beta n(m^2,J) \\&=
\frac{a-\JJ^2}{m^6} + \frac{\alpha}{m^4} + \beta \frac{\JJ^4-b\JJ^2}{m^8}\,,
\end{aligned}\label{warm up F} \ee
where we must find $\alpha,\beta$ such that $F(m^2,J)\geq 0$ for all $J=0,2,4\ldots$ and $m\geq M$.
Taking the average of any such $F$ then proves $h_3 \geq -\alpha h_2$.
The optimal bound will come from a non-negative $F$ with minimal $\alpha$.

Let us first reproduce the first Cauchy-Schwarz argument in this language, which should give
$\alpha=\frac{b}{M^2}$.
Assume $\beta>0$.  The argument amounts to completing squares in the $\JJ^4/m^8$ term:
\be
 F(m^2,J) =\frac{\alpha-\beta \lambda^2}{m^4}+ \frac{a}{m^6} + \frac{\JJ^2}{m^6}\left(2\beta \lambda-1\right) - \beta b\frac{\JJ^2}{m^8}
  + \underbrace{\beta \frac{1}{m^4}\left(\frac{\JJ^2}{m^2} -\lambda\right)^2}_{\rm positive}\,.
\ee
For any $\lambda$ this is an identical rewriting of eq.~\eqref{warm up F}, and the Cauchy-Schwartz-like method is to
choose $\lambda,\beta$ such that the other terms are positive as well.
From the limit $\JJ\to\infty$, the terms with $\JJ^2$ need to give a positive functions of $m$,
which imposes that $(2\beta \lambda-1- \beta b/M^2)\geq 0$.
To minimize $\alpha$ we must minimize $\beta \lambda^2$;
we find that the minimum saturates the inequality, and is simply $\beta=M^2/b$ with $\beta \lambda=1$.
With this choice, our trial functional becomes
\be
F(m^2,J) =\frac{\alpha-\frac{b}{M^2}}{m^4} + \frac{a}{m^6}+
\underbrace{\frac{\JJ^2}{m^6} \left(\frac{1}{M^2}-\frac{1}{m^2}\right)
+\frac{M^2}{b m^4}\left(\frac{\JJ^2}{m^2} -\frac{b}{M^2}\right)^2}_{\rm positive}\,.
\label{F with pos}
\ee
The minimal $\alpha$ for which the first terms are positive for all $m^2$
is then $\alpha=\frac{b}{M^2}$, precisely as anticipated!
We have thus exhibited a positive functional $F$ which proves the Cauchy-Schwarz bound in \eqref{warm up h3 bound}.

It is now easy to see why this bound is not optimal: $F$ doesn't \emph{need} to be expressible as a sum
of three separately positive parts!

In $d=4$, for example, the above argument gives
$-16 \leq \tilde{g}_3$. In comparison, using the numerical search strategy detailed in the next subsections,
we find that the following combination is positive for all $J=0,2,4\ldots$ and $m\geq M$:
\begin{align}
0 \leq F(m^2,J) = \begin{pmatrix}
10.61249\\ 
1 \\ 
0.0671875
\end{pmatrix} \cdot \begin{pmatrix}
g_2(m^2,J) \\ 
M^2g_3(m^2,J) \\ 
M^4n_4(m^2,J)
\end{pmatrix}. \label{optimal g3}
\end{align}
This allows to infer, by taking the average of this inequality, that
\begin{align}
-10.61249 \leq \tilde{g}_3\,.
\end{align}
This is significantly stronger than $-16\leq \tilde{g}_3$ that we just derived in an ad hoc manner.  How can we understand the solution \eqref{optimal g3} analytically? They key ingredient will be that spins $J$ are discrete,
whereas our ad hoc bound treated $J$'s as continuous parameter.

\begin{figure}[t]
        \centering
        \includegraphics[width=0.65\linewidth]{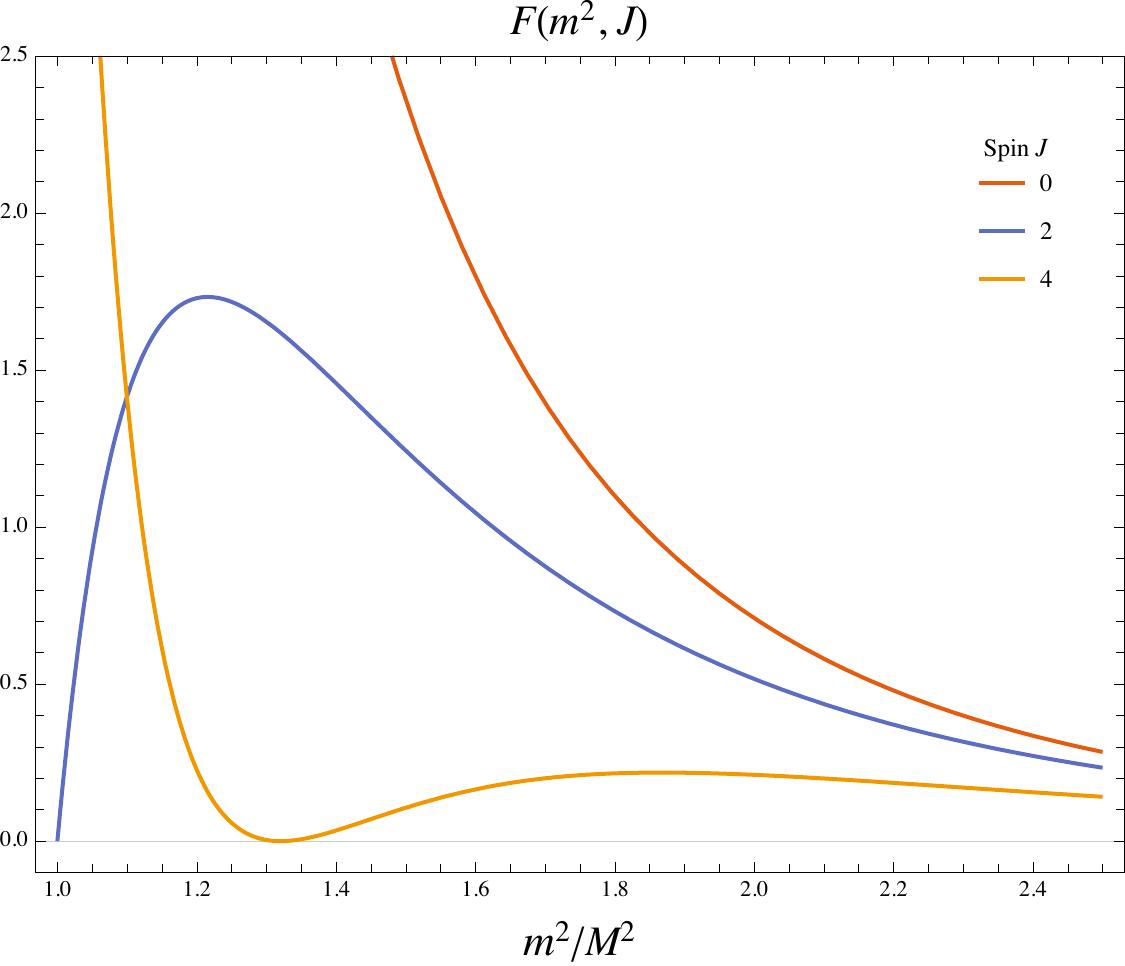}
        \caption{\label{fig:4d_spin_functional}
The non-negative function $F(m^2,J)$ returned by \texttt{SDPB} when maximizing the lower bound for $\tilde{g}_3$. The spin 4 curve is particularly enlightening because it features a double root.  
}
\end{figure}

We now calculate this bound analytically. Let us return to the functional ansatz \eqref{warm up F} and try to directly constrain the unknowns $\alpha$, $\beta$.
Putting $J=0$ we only deduce $\alpha\geq 0$. Putting $J=2,4\ldots$ gives a sequence of
quadratic polynomials in $(m^2-M^2)$, each with positive curvature. Such polynomials are non-negative if the two roots are negative,
\emph{or} if both roots are positive and equal, \emph{or} if they are complex conjugate pairs. It seems virtually
impossible to guess a priori which case is realized, however this information is readily
gleaned by plotting the numerical polynomials \eqref{optimal g3},
as shown in fig.~\ref{fig:4d_spin_functional}. We see that the $J=2,4$ inequalities are both saturated:
the former by having a root at $m^2=M^2$, and the latter by having a positive double root (i.e. vanishing discriminant).

These two saturated inequalities give algebraic equations that may be solved analytically;
this determines the vector $(\alpha_*,1,\beta_*)$ in eq.~\eqref{optimal g3} to be:
\be
\alpha_*= \frac{5d-2+\sqrt{\frac{(d+3)(319d^3+76d^2-292d+32)}{6(d+1)(d+4)}}}{2(d-2)},\quad \beta_* = \frac{1}{2d(d-1)}
\left( \alpha_*-\frac{5d-2}{d-2}\right)\,.
\ee
In $d=4$, this gives $\alpha_*=\frac92+\frac74\sqrt{\frac{61}{5}}\approx 10.61249\ldots$ precisely as found numerically!
This result showcases the use of numerics to guide our analytical understanding.
Moreover, the preceding formula can be shown to give the correct optimal lower bound on $\tilde{g}_3$ in any $d\geq 3$,
when including the single null constraint $n_4$.
(When we add more equations below, the $d=4$ bound will converge to $\tilde{g}_3 \geq -10.346$.)

\subsection{Dual problem: general formulation}
\label{ssec:dual}

We now introduce the general ``dual'' optimization problem which allows
to carve out the space of EFT coefficients $\tilde{g}_3, \tilde{g}_4, \tilde{g}_5, \dots$ allowed by unitarity and positivity.  

The data at our disposal comes from the $B_k$ sum-rules in eq.~\eqref{Bks}:
\begin{enumerate}
	\item Representative averages $g_k(m^2,J)$ which measure each ``desired'' $g_k$:
	\begin{align}
		g_2 = \avg{g_2(m^2,J)}, \quad g_3 = \avg{g_3(m^2,J)}, \quad g_4 = \avg{g_4(m^2,J)},\quad \cdots
	\end{align}
	\item A set of null functions $n_i(m^2, J)$ whose heavy averages vanish:
	\begin{align}
		0 = \avg{n_4(m^2,J)} = \avg{n_5(m^2,J)} = \avg{n_6(m^2,J)} = \dots
	\end{align}
\end{enumerate}
First consider the problem of lower-bounding $g_3$, given $g_2$, but being agnostic about the other $g_k$'s with $k\geq 4$.
Define a vector of functions which combines $g_2,g_3$ and null constraints:
\be
 v(m^2,J) \equiv \big( g_2(m^2,J),\  M^2 g_3(m^2,J),\ n_4(m^2,J),\ n_5(m^2,J),\ \ldots\big)\,. \label{def v}
\ee
The following (``dual'') optimization problem then determines a lower bound on $g_3$:
\begin{align}\left\{
\begin{array}{ll}\displaystyle
	\text{maximize:} & \quad A\\\displaystyle
	\text{subject to:} & \quad 0 \leq \big({-}A,1,c_4,c_5,\ldots\big)\cdot v(m^2,J)	\qquad \forall\ m\geq M,\  \forall J=0,2,4,\ldots
\end{array}
\right.
\label{eqn:lower_bound}\end{align}
Here we are maximizing over $A$  and all possible linear combinations $c_i$ of the null constraints $n_i(m^2,J)$.
Having found such a linear combination $c_i$ and an optimal number $A$,
it readily follows from linearity and positivity of $\avg{\cdot}$ that
\begin{align}
	0 \leq -Ag_2 + M^2 g_3 + 0,
\end{align}
allowing the desired \emph{optimal} bound on $\tilde{g}_3$ to be inferred:
\begin{align}
	A \leq \tilde{g}_3.
\end{align}
This equation motivates the maximization of $A$. Alternatively, to find an optimal \emph{upper} bound $B$,
it is enough to consider the analogous problem:
\begin{align}
\left\{\begin{array}{ll}\displaystyle
	\text{minimize:} & \quad B\\\displaystyle
	\text{subject to:} & \quad 0 \leq \big(B,{-}1,c_4,c_5,\ldots\big)\cdot v(m^2,J)	\qquad \forall\ m\geq M,\  \forall J=0,2,4,\ldots
\end{array}
\right.\label{eqn:upper_bound}\end{align}
It then follows that $\tilde{g}_3\leq B$.
Taking the intersection of these two sets yields a convex region defined by 
\begin{align}
	A \leq \tilde{g}_3 \leq B.
\end{align}
These are the inequalities inferred on $\tilde{g}_3$ from data in the $B_k$ sum-rules.

In theory, the null constraints $n_i$ are part of an infinite dimensional vector space, but for numerical purposes, the dimensionality is taken to be finite and determined by the truncation order of the low-energy expansion $\mathcal{M}_{\text{low}}$.
We will find that the optimal bounds converge rapidly as the maximal degree is increased.

This problem can be readily adapted if we are given additional assumptions.
For example, to make exclusion plots in the $(\tilde{g}_3,\tilde{g}_4)$ plane, one strategy is to
postulate some value of $\tilde{g}_3^{(0)}$ in the allowed range,
and for each value we determine lower and upper bounds on $\tilde{g}_4$.
The vector $v$ in eq.~\eqref{def v} then acquires an extra $g_4$ row:
\be
v \mapsto
 \big( g_2,\  M^2 g_3,\ M^4g_4,\ n_4,\ n_5,\ \ldots\big)\,. \label{def v 4}
\ee
Imposing positivity of $\big({-}A,{-}B,1,c_4,c_5,\ldots\big)\cdot v(m,J)$ for all $m$ and $J$
yields a lower bound $\tilde{g}_4\geq A+B\tilde{g}^{(0)}_3$, for example.
In fact, since the EFT parameters enter linearly, the resulting inequality automatically
carves out a half-space in the $(\tilde{g}_3,\tilde{g}_4)$-plane:
\be
\tilde{g}_4 - B\tilde{g}_3 \geq A.
\ee
This half-plane is tangent to the allowed region at $\tilde{g}_3=\tilde{g}_3^{(0)}$.
If this process is repeated for distinct $\tilde{g}_3^{(0)}$, a collection of planes is rapidly generated
from which it is possible to carve out the convex allowed space.

This processes generalizes to higher dimensional planes (i.e. hyperplanes):
the vector $v$ always contains only those coefficients we are not being agnostic about,
plus an arbitrary number of null constraints. Although we will not go beyond three-dimensional regions,
we note that an efficient search algorithm in higher dimensions is described in ref.~\cite{Chester:2019ifh}.

\subsection{Example with a \texttt{SDPB} implementation}
\label{ssec:sdpb}

The optimization problem just formulated is in a form that is directly amenable to the \texttt{SDPB} solver \cite{Simmons-Duffin:2015qma}.
There are just two simple substitutions to make:
\begin{enumerate}
	\item The program accepts polynomials of $x\geq 0$; we set $m^2 \mapsto M^2(1+x)$ and remove a common denominator.
	\item The program  accepts finite lists of polynomial constraints. We tabulate a finite list of spins
	 $J = 0,2,\dots,J_{\text{max}}$ and add a single function of $x$ corresponding to $J\to\infty$.
\end{enumerate}
The second truncation is valid as long as $J_{\text{max}}$ is taken sufficiently large; once convergence is achieved, further increases of $J_{\rm max}$ have no effect on the bounds.


We consider now an example relevant to one of the plots in the next section,
when working in $d=4$, $J_{\text{max}}=40$ and Mandelstam order $n=4$.
The goal is to find a lower bounding plane on $\tilde{g}_4$ for
fixed $\tilde{g}_3$, say $\tilde{g}_3 = -10.5$ (which is slightly above the allowed lower bound \eqref{optimal g3} for this truncation).
The polynomial vector $v$ combines the three observed coefficients and a single null constraint $n_4$,
rescaled by appropriate powers of $M$:
\begin{eqnarray}
v(x,J)
&\equiv& M^4(1+x)^4\times \left( g_2,M^2 g_3,M^4g_4,M^4n_4 \right) \\
 &=& \left( (1+x)^2,  (1+x)(3-2J(J+1)),\ \tfrac{1}{2}, 2J(J+1)(J(J+1)-8)\right).
\end{eqnarray}
In addition to tabulating $v$ for all even $J\leq J_{\rm max}$, we also include the infinite-$J$ limit,
which is simply the coefficient of $J^4$:
\be
 v(x,\infty) =\left( 0,\ 0,\ 0,\ 2\right)\,.
\ee
To lower-bound $\tilde{g}_4$ at the stated value of $\tilde{g}_3$,
we search for four-vectors $y$, normalized to $y{\cdot}\big(0,0,1,0\big)=1$,
which solve the following problem:
\begin{align}
\left\{\begin{array}{ll}\displaystyle
	\text{maximize:} & \quad h=y{\cdot}\left( -1,-(-10.5),0,0\right)\\\displaystyle
	\text{subject to:} & \quad 0 \leq y{\cdot} v(x,J) \qquad \forall\ x>0, J\in\{ 0,2,\cdots,40,\infty\}\,.
\end{array}
\right.\end{align}
The lower bound is then
\be
 \tilde{g}_4\geq h \qquad (\mbox{at } \tilde{g}_3=-10.5)\,.
\ee
The solution vector $y$ computed by \texttt{SDPB} was found to be
$y\approx \left(1.4823,\ 0.1810,\ 1,\ 0.01472\right)$, giving $h=0.4183$. The half-plane allowed by positivity of $y{\cdot}v$ is thus
\be
 \tilde{g}_4 + 0.1810(\tilde{g}_3+10.5) \geq 0.4183
\ee
which gives one of the boundaries used to make the $n=4$ region in fig.~\ref{fig:kinks} (b) below.

\subsection{Generating null constraints}

So far we used a single constraint from crossing: the null average $n_4(m^2,J)$ from eq.~\eqref{eqn:null1}.
The bounds improve after we add more constraints.  Let us describe a way to generate them.

A straightforward method is as follows.
First, fix a degree in Mandelstam invariants, $n$.
Then, list all the $\left \lfloor\tfrac{n}{2}\right \rfloor$ low-energy averages corresponding to this degree, namely
the coefficient of $t^{n-k}$ in the left-hand-side of the $B_k(t)$ sum rules \eqref{Bks} with $k\leq n$,
using the crossing-symmetric low-energy ansatz \eqref{eqn:eft_amplitude}.
The right-hand-sides of those linear combinations with vanishing left-hand-side then constitute a basis of null constraints.

The first few cases, up to degree $n=7$, are (in arbitrary normalization):
\be\begin{aligned}\label{null constraints}
m^8 n_4 = &\  (4-5 d) \JJ^2+2\JJ^4\,,\\
m^{10} n_5 = & \left(23 d^2-12 d-20\right) \JJ^2+(-21 d-2) \JJ^4+4\JJ^6\,,\\
m^{12} n_6 = &\ 4 \left(13 d^2+21 d+2\right) \JJ^4-3 (d+2) \left(17 d^2+4 d-16\right) \JJ^2+2 (-9 d-8) \JJ^6+2\JJ^8\,,\\
m^{14}n_7 = &\ 6\left(45 d^2+140 d+92\right) \JJ^6+4 \left(-140 d^3-619 d^2-711 d-46\right) \JJ^4
\\&+\left(401 d^4+2108 d^3+2284 d^2-3008 d-3360\right) \JJ^2+5 (-11 d-18) \JJ^8+4\JJ^{10}\,,\\
m^{14}n_7^\prime = & \left(-27 d-14\right) \JJ^4+2(d-1) (19 d+22) \JJ^2+4\JJ^6
\end{aligned}\ee
where we recall that $\JJ^2=J(J+d-3)$. Note that \emph{all} null averages vanish when $J=0$: they relate spinning heavy states to one another, but spinless states
are completely decoupled.

We find that there exists a single null constraint for each degree $n=4,5,6$, then two constraint each for $n=7,8,9$, three each for $n=10,11,12$, etc.:
the number of linearly independent null constraints at each degree increases by 1 for every increase of $n$ by 3.
A sequence of generating functions $X_k(t)$ which
enumerates them all is discussed in appendix \ref{app:Xn}.

Finally, it is important to stress that the null constraints only average to zero modulo EFT loops,
since the method for finding them relies on the explicit tree-level parameterization \eqref{eqn:eft_amplitude}.
The interpretation of resulting inequalities as bounds on $g_k$ is thus only strictly valid in this approximation.
In an interacting EFT, the coefficients $g_k$ depend on choices of scale and renormalization scheme,
and the correct interpretation of the positive functionals $F$ is that they give rigorous
(possibly non-optimal) inequalities of the form:
\be
\avg{g_3(m^2,J)+ \sum_i c_i n_i(m^2,J)} \geq A \avg{g_2(m^2,J)} \label{scheme indep}
\ee
where all averages are computed as integrals over arcs with $|s|\approx M^2$ following eq.~\eqref{Bks general}.
The method thus produces rigorous bounds on computable combinations of EFT couplings at the scale $M$.

\subsection{An ad hoc upper bound on $(\partial\phi)^4$}
\label{ssec:g2}

The systematic method explained above bounds ratios $g_k/g_2$, but how about $g_2$ itself?
Here we present one upper bound on $g_2$;
this subsection is somewhat separate from the rest since we were unable to systematically optimize the bound.

The naive intuition is that if heavy states couplings are order unity, then $g_2\sim \frac{1}{M^d}$.
An upper bound with this scaling should thus follow from the unitarity limit, $\rho_J\leq 2$.
This would be the full story \emph{if} heavy states only had a finite number of spins, however,
to get an actual bound one must also control the infinite sum over spins.
The idea is to combine the sum rule for $g_2$ with a multiple of the first null constraint $n_4$:
\be \label{g2strong average}
g_2 = \avg{\frac{1}{m^4} - \alpha \frac{f(J)}{m^8}} \quad\mbox{with}\quad f(J)= \JJ^2\big(2\JJ^2-(5d-4)\big)\,.
\ee
This holds for any $\alpha$; we recall that $\JJ^2=J(J+d-3)$ is the spin Casimir.
Inserting the definition \eqref{avg def} of the average and switching to a dimensionless mass parameter $z=M^2/m^2<1$, this can be rewritten
\be
g_2= \frac{1}{\pi M^{d}} \sum_{J \text { even }} n_{J}^{(d)} \int_{0}^{1} \frac{dz}{z}z^{\frac{d}{2}} \prob_J(M^2/z)\left(1 - \alpha z^2 f(J)\right)\,.
\ee
This is a good step
to get an upper bound on $g_2$ since for $\alpha>0$ the integrand is mostly negative at large spin $J$, except for a small region
with $z$ small. We use unitarity in two steps: first, we use positivity $\rho_J\geq 0$ to restrict the integral to the small-$z$ region
where the parenthesis is positive, where we can then use $\rho_J\leq 2$. Thus:
\be
 g_2 \leq \frac{2}{\pi M^{d}}\sum_{J \text { even }} n_{J}^{(d)} \int_0^{z_{\rm max}(J)} \frac{dz}{z} z^{\frac{d}{2}} \left(1 - \alpha z^2f(J)\right)
\ee
where $z_{\rm max}(J)= \min([\alpha f(J)]^{-\frac12},1)$ determines the region where the parenthesis is positive.
The important point is that at large spin this region shrinks, which will ensure convergence of the $J$ sum.
Effectively the region is bounded by impact parameter $M\tilde{b}\lsim \alpha^{-1/4}$.
At small spins the full range is generally accessible.
Letting $z_{\max}(J)=1$ for $J\leq J_*(\alpha)$, the sum splits as
\be
 g_2\leq \frac{2}{\pi M^{d}} \left[
 \sum_{0\leq J \leq J_*(\alpha)} n_J^{(d)} \left( \frac{2}{d} - \frac{2\alpha f(J)}{d+4}\right)
 + \frac{8 \alpha^{-\frac{d}{4}}}{d(d+4)}\sum_{J > J_*(\alpha)} \frac{n_J^{(d)}}{f(J)^{\frac{d}{4}}}\right]\,.
\ee
Both sums run only over even spins. Since $n_J^{(d)}\propto J^{d-3}$ at large spin, the sum to infinity converges.
This inequality is valid for any $\alpha>0$; with increasing $\alpha$, the $J=0,2$ terms tend to increase whereas the rest decreases:
the optimal bound with this method is obtained by minimizing over $\alpha$.
The dependence is rather non-linear (which is why we were not able to generalize the method to include more null constraints),
but evaluating the sum numerically we find (in all dimensions we tried) that the optimum occurs with $J_*=2$.
In $d=4$, for example, the optimal value $\alpha_*=0.025$, giving the analytic bound:
\be
 0\leq \frac{g_2}{(4\pi)^2} \leq \frac{0.794}{M^4} \qquad(d=4)\,. \label{g2 upper}
\ee
As expected, up to a standard loop factor, the coefficient of $\frac12(\partial\phi)^4$ can't exceed order unity in units of the heavy scale.  We stress that, contrary to other bounds in this paper, upper bounds on couplings at the cutoff scale
cannot be straightforwardly interpreted in terms of Lagrangian parameters, since any EFT which saturates them is by definition strongly interacting already below the cutoff, making quantum corrections non-negligible. Rather, the bound may be interpreted as follows: among all observables which are linear in the $S$-matrix at the scale $M$ and which reduce to $g_2$ at weak coupling, there exists one which satisfies eq.~\eqref{g2 upper}: namely, eq.~\eqref{g2strong average} with the quoted $\alpha_*$.
This observable resides at the scale $M$ in the sense of eq.~\eqref{scheme indep}. 

Similar bounds in different spacetime dimensions are recorded in table \ref{tab:g2 upper}; after normalizing with a suitable $d$-dimension factor (closely related to $n^{(d)}_0$) we find that the obtained bound is pretty stable in dimensions.

\begin{table}[t]
\centering
\begin{tabular}{c|cccccccccc}
\hline
$d$ & 3 & 4 & 5 & 6 & 7&8 & 9 & 10 & 26 & 50\\ \hline
bound & 0.155 & 0.132 & 0.138 & 0.140 & 0.141 & 0.141 & 0.140 & 0.140 & 0.132 & 0.127 \\\hline
\end{tabular}
\caption{Nonperturbative (possibly non-optimal)
upper bounds on $\frac{g_2 M^d\Gamma(\tfrac{d}{2})}{(4\pi)^{\frac{d}{2}}\Gamma(d)}$ in various spacetime dimensions.
\label{tab:g2 upper}
}
\end{table}


\section{Numerically ruling-out: The allowed space of scalar EFTs}

\label{sect:numerics}

In this section, we summarize the obtained numerical results, focusing on $d=3+1$
space-time dimensions. 
Treating the low-energy EFT to tree-level, we determine the space of EFT coefficients $g_n$, where the low-energy amplitude is parameterized as
\begin{align}\begin{split}
\mathcal{M}_{\text{low}} = & -g^2 \left[\frac{1}{s} + \frac{1}{t} + \frac{1}{u}\right] - \lambda\\
& + g_2(s^2 + t^2 + u^2) +g_3 (stu) + g_4(s^2 + t^2 + u^2)^2 + g_5(s^2 + t^2 + u^2)(stu)+  \cdots
\end{split}\end{align}
Recall that it is convenient to introduce dimensionless EFT coefficients $\tilde{g}_n$ normalized by $g_2$ and appropriate powers of the mass gap $M$ introduced in equation \eqref{eqn:normalized_couplings}, since the numerical analysis is performed directly on these variables.
We find optimal upper and lower bounds for $\tilde{g}_3$, $\tilde{g}_4$, and $\tilde{g}_5$, given positivity of high-energy spectral densities $\rho_J$,
using the optimization framework introduced above.

\begin{table}[t]
\centering
\begin{tabular}{|c|c|c|c|c|}
\hline
$n$ & $\text{dim}\ N$     & $\min\tilde{g}_3$ & $\min \tilde{g}_5$ &  $\min \tilde{g}_6^\prime$  \\\hline
4   & 1                         & -10.6125      & $-\tfrac{25}{6}\approx -4.17$  & -15  \\ \hline
5   & 2                        & -10.6125       & $-\tfrac{25}{6}\approx -4.17$   &    \\ \hline
6   & 3                        & -10.3662        & -4.0969  & -13.436  \\ \hline
7   & 5                        & -10.3580          &   & \\ \hline
8   & 7                        & -10.3492          &  -4.0960& -12.926\\\hline
9   & 9                        & -10.3492           &  & \\ \hline
10  & 12                     & -10.3477           &  & \\ \hline
11  & 15                     & -10.3474         &     & \\ \hline
12  & 18                     & -10.3473         &    -4.0960& -12.820 \\\hline
13  & 22                     & -10.3470         &    &\\ \hline
14  & 26                     & -10.3468         &     &\\ \hline
15  & 30                     & -10.3466         &    & \\ \hline
16  & 35                    & -10.3465           & -4.0960  & -12.811\\ \hline
\end{tabular}
\caption{Convergence of the ``box bounds'' \eqref{eqn:box_bounds} with increasing Mandelstam order $n$ , which results in larger sets
of orthogonal constraints $N=\{n_4(m^2,J), \cdots\}$. At order $n=3$, all lower bounds would be $-\infty$ since no null constraints exist at that order.  
}
\label{tab:convergence_table}
\end{table}

\subsection{Bounds on individual coefficients}

Let us begin by confining the value of individual EFT coefficients, being completely agnostic about all the others.
Table \ref{tab:convergence_table} shows how these bounds depend on the number of crossing symmetry constraints
kept. We conclude that, in $d = 3+1$ space-time, normalized EFT coefficients satisfy:
\begin{align}
-10.346 \leq \tilde{g}_3 \leq 3, \qquad
0\leq \tilde{g}_4 \leq \frac{1}{2}, \qquad
-4.0960 \leq \tilde{g}_5 \leq \frac{5}{2}\,.
\label{eqn:box_bounds}
\end{align}
The simple, rational upper bounds are saturated by the spin-0 contribution to sum rules like eq.~\eqref{eq:list_of_gs}.

While keeping constraints with higher Mandelstam degree $n$ is feasible, runtime-wise, we found that the large-spin convergence was harder to control as
we needed $J_{\text{max}} = \mathcal{O}(1000)$.
However, for $n \leq 16$, convergence is easily obtained for smaller spins. This is the reason why we stopped the table at $n=16$;
it would be interesting to understand how to stabilize the numerics at large $n$.

\begin{figure}[t]
        \centering
        \includegraphics[width=0.65\linewidth]{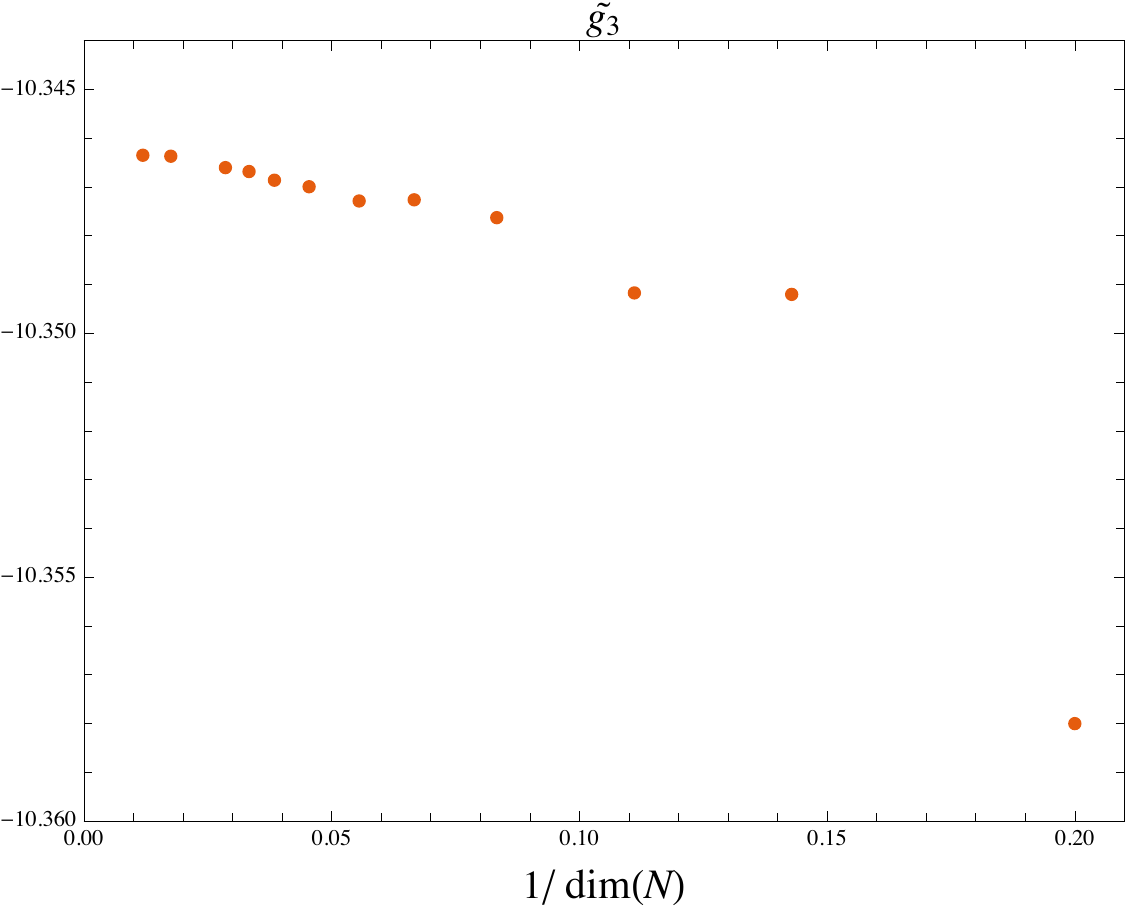}
        \caption{\label{fig:convergence}
The $\tilde{g}_3$ lower bound vs. $1/ \text{dim}\,N$, where $\text{dim}\,N$ is the dimension of the number of null constraints accessed.
Plotting with respect to $1/({\rm dim}\ N)^2$ or different powers of $1/n$ did not reveal any particularly compelling fitting function.
The kink described in the next plot is situated at $-10.19 \approx \tilde{g}_3$.
}
\end{figure}

The $\tilde{g}_3$ lower bound is plotted as a function of $1/ \text{dim}\,N$ in fig.~\ref{fig:convergence}, where $\text{dim}\,N$ is the dimension of the number of null constraints accessed.  The dimension of this vector space is naturally monotonically increasing with the Mandelstam order $n$.
Ideally, one would like to extrapolate the bounds to $n\to\infty$, however
the approach is somewhat irregular and we didn't find a compelling fit function. Therefore, the recorded bounds are simply taken from our largest reliable value of $n$. 

In $d=6$ and $d=10$, the analogous bounds take the values:

\begin{align}
d=6: & \quad
-8.13879\leq \tilde{g}_3 \leq 3, \qquad
0\leq \tilde{g}_4 \leq \tfrac{1}{2}, \qquad
-3.03373 \leq \tilde{g}_5 \leq \tfrac{5}{2}\\
d=10: & \quad
-7.04934\leq \tilde{g}_3 \leq 3, \qquad
0\leq \tilde{g}_4 \leq \tfrac{1}{2}, \qquad
-2.51054 \leq \tilde{g}_5 \leq \tfrac{5}{2}
\end{align}


These upper and lower bounds were obtained by solving the optimization problems given in equations \eqref{eqn:lower_bound} and \eqref{eqn:upper_bound}, respectively.
For the $n=16$ data points in the table, we kept spins up to 200 (in addition to $J=\infty$).
A full list of the bounds on the first $10$ coefficients $\tilde{g}_n$ obtained with $n=10$ (corresponding to 12 linearly independent null constraints)
are presented in appendix \ref{app:operator}.

\subsection{Two-dimensional allowed region in $(\tilde{g}_3,\tilde{g}_4)$ plane}

\begin{figure}[t]
    \centering
    \includegraphics[width=0.7\textwidth]{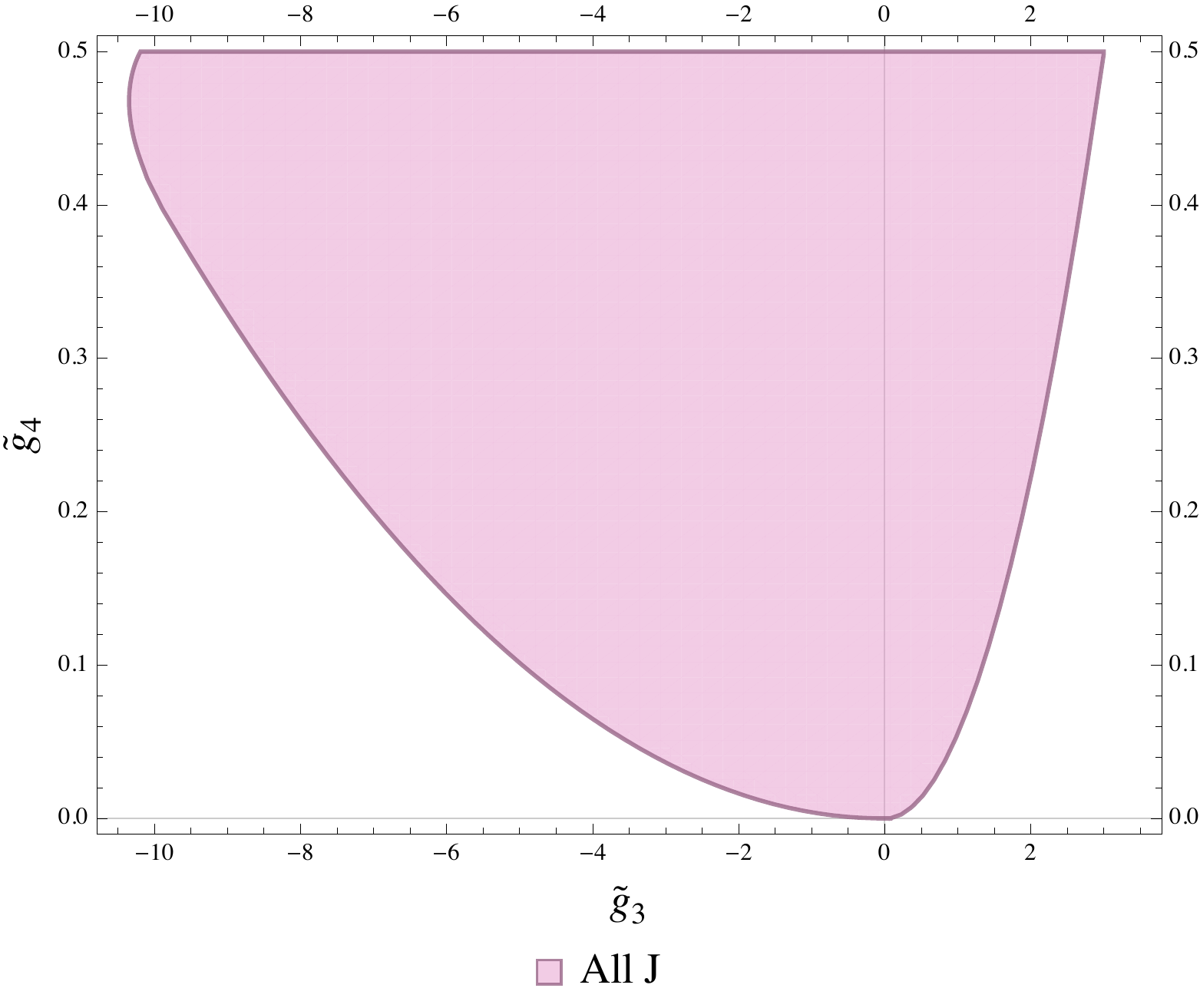}
    \caption{The $(\tilde{g}_3, \tilde{g}_4)$ allowed region.  Numerics were performed at $n=10$ Mandelstam order and $J=0,2,\dots,40$.  One can see that $g_3$ may take-on negative values, while $g_4$ is positive.  Boundaries appear smooth except for two kinks at $(-10.19,0.5)$ and $(3,0.5)$.}
    \label{fig:exclusion}
\end{figure}

The ``box bounds'' in eq.~\eqref{eqn:box_bounds} tell an incomplete story since they miss potential correlations between the coefficients.
Exclusion plots on the $(\tilde{g}_3,\tilde{g}_4)$ plane can be obtained following the constrained optimization framework presented at the end of
section \ref{ssec:dual}.  Specifically, upper and lower bounding planes on $\tilde{g}_4$ were obtained by sampling $\tilde{g}_3$ at various points in the range allowed by
eq.~\eqref{eqn:box_bounds}. Repeating this process for a large number of sampling points generates a collection of linear inequalities, whose intersection
defines a refined allowed region. Figure \ref{fig:exclusion} depicts this region with over $100$ sampling points along $\tilde{g}_3$.

Further insight into the shape of the region can be obtained by noticing that the crossing symmetry constraints
do not mix particles of spin 0 with the others, as noted in eq.~\eqref{null constraints}.
This indicates that heavy spin-0 particles satisfy crossing on their own, as will be further discussed in the next section.
For numerical purposes, this decoupling allows to refine the problem: any high-energy spectrum can be written as a positive sum of
its spin-0 content, plus a unitary solution to crossing that only contains particles of spin $J\geq 2$.
The full allowed region is then simply the convex hull of the allowed regions for these two problems:
\begin{align}
	\text{Entire region} = \text{Convex Hull}\left[\text{Spin-0} + \text{Spin-}J\geq2\right].
\end{align}
As may be seen from the form of the $g_3$ sum rule \eqref{eq:list_of_gs},
the two solutions are differentiated by the sign of $g_3$: positive for Spin-0 and negative for Spin-$J\geq 2$.

\begin{figure}[t]
    \centering
    \includegraphics[width=0.7\textwidth]{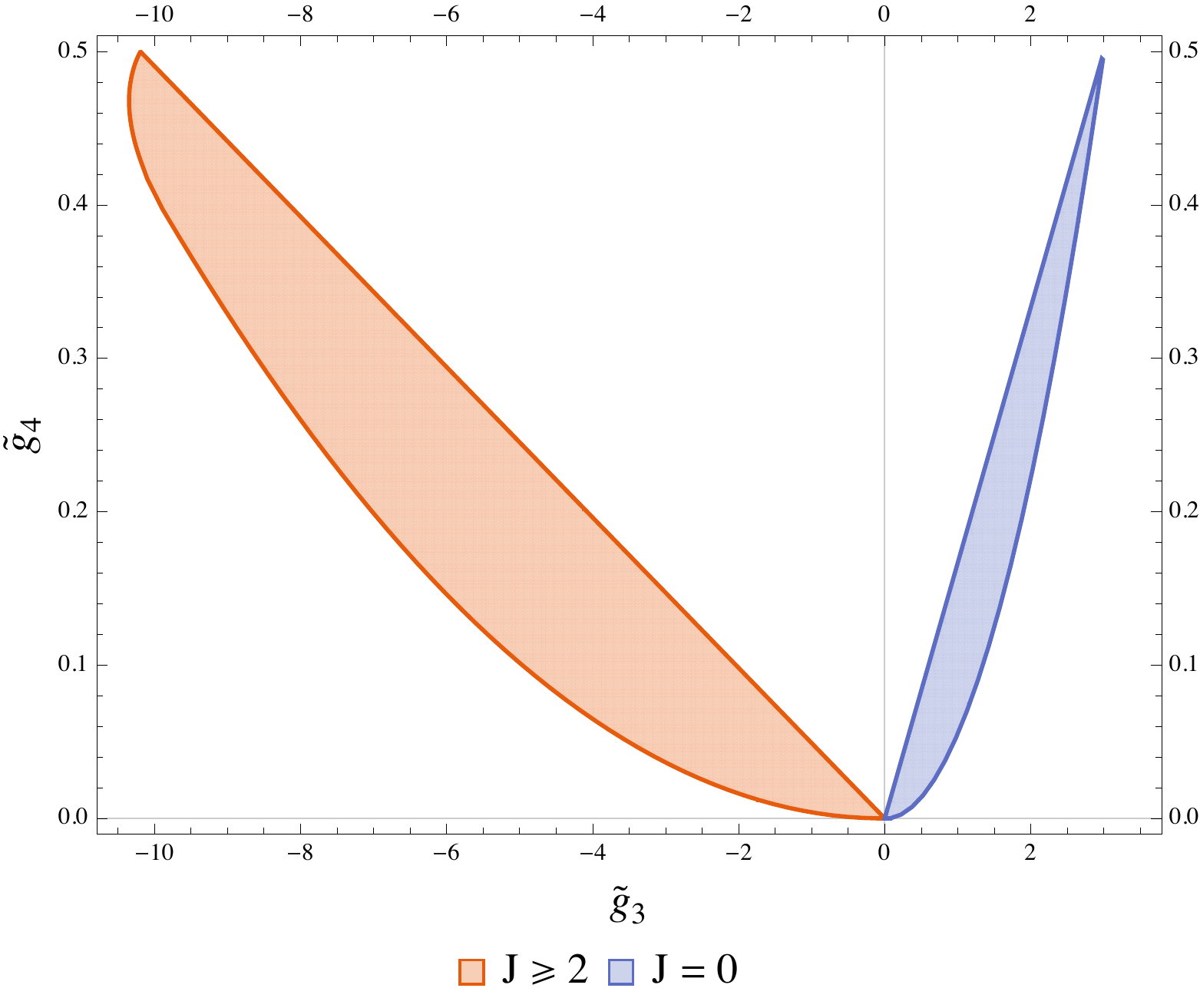}
    \caption{The $(\tilde{g}_3, \tilde{g}_4)$ allowed region, segmented into theories without spinless particles $(J\geq 2$) and theories with only spinless particles $(J=0)$. 
    The convex hull of these two regions reproduces fig.~\ref{fig:exclusion}.}
    \label{fig:decomposition}
    \end{figure}

In our implementation of the dual problem, theories with only $J\geq 2$ particles can be studied by simply dropping the positivity constraint for the
functional action on $J=0$. The allowed regions for the Spin-0 and Spin-$J\geq2$ sub-problems are the narrow almond-shaped regions shown
in fig.~\ref{fig:decomposition}.

The shape of these regions is largely explained by a simple scaling argument:
given any solution to crossing, scaling-up its overall mass scale will give a new solution.
Starting from any allowed point $(\tilde{g}_3,\tilde{g}_4)$, this generates an allowed path $(\alpha \tilde{g}_3,\alpha^2\tilde{g}_4)$
where $0\leq \alpha\leq 1$.  This explains the parabolic shape of the ``underbellies'' in fig.~\ref{fig:decomposition}. 
In fact the Spin-0 almond is simply the convex hull of the parabola connecting $(0,0)$ to $(3,\frac12)$.
(This is qualitatively similar to what is found in the forward limit \cite{talks,Bellazzini:2020cot}.)

\begin{figure}
    \begin{subfigure}[t]{0.49\textwidth}
        \centering
        \includegraphics[width=1\linewidth]{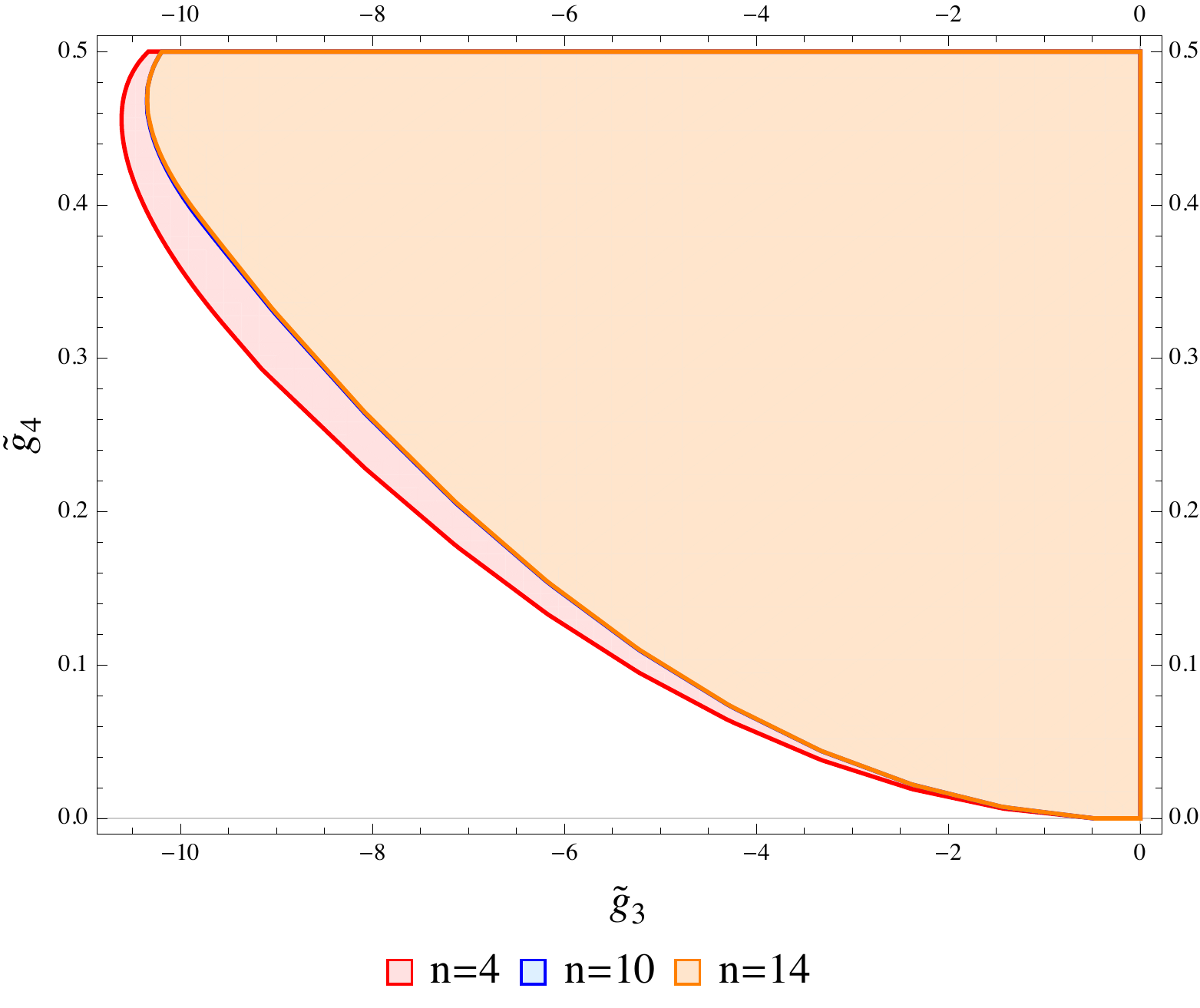}
      \caption{Convergence is rapidly achieved by adding null constraints.
      }
    \end{subfigure} 
    \begin{subfigure}[t]{0.49\textwidth}
        \centering
  		\includegraphics[width=\linewidth]{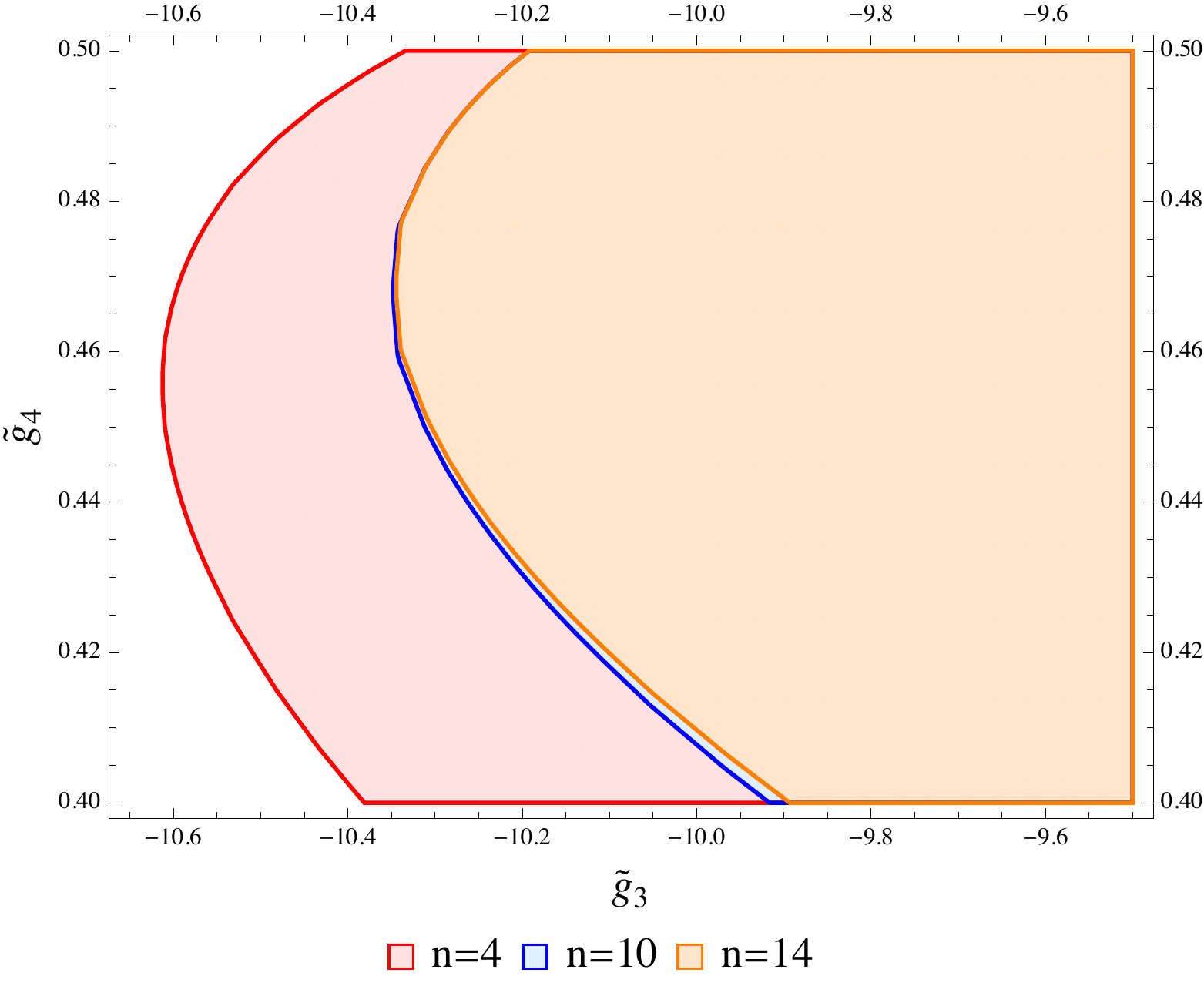}
  		\caption{Close-up near the left kink at $(-10.19,0.5)$.}
    \end{subfigure}
        \caption{
Convergence of the $(\tilde{g}_3, \tilde{g}_4)$ allowed region (shaded) with the Mandelstam degree $n$ of crossing equations.
The positive section of the horizontal axis is omitted since that region converges immediately.}
\label{fig:kinks}
\end{figure}

The Spin-$J\geq 2$ region is more complicated -- while it also displays a parabolic underbelly near the origin, it fails to extend
all the way to $\tilde{g}_4=\frac12$. The boundary must thus exhibit non-analytic behaviour at the end of the parabola,
however we were unable to localize a kink that remains stable with varying the Mandelstam expansion order $n$,
suggesting milder non-analyticity (such as a discontinuous second derivative). This qualitative feature is demonstrated by close-ups on the high-$\tilde{g}_4$ end of the allowed region are shown in fig.~\ref{fig:kinks}, which also indicates convergence with increasing $n$.

The line $\tilde{g}_4=\frac12$ features two kinks: at $(-10.19, 0.5)$ and $(3,0.5)$, as shown in figure \ref{fig:kinks}.
Below we will find analytic expressions for the scattering amplitude at these kinks!

We chose the simplest and perhaps weakest method of calculating the exclusion region.  Method of radials or normals have the potential more efficiently calculate these boundaries with higher fidelity near kinks \cite{Cordova:2019lot}. For example, the normals maximization procedure explained extracts sharp features at regions of large curvature.  Unfortunately, we were not able to use the normals method because it requires one to solve the ``primal" problem instead. 

\subsection{Three-dimensional allowed volume in $(\tilde{g}_3,\tilde{g}_4,\tilde{g}_5)$}

\begin{figure}
    \begin{subfigure}[t]{0.56\textwidth}
        \centering
        \includegraphics[width=1\linewidth]{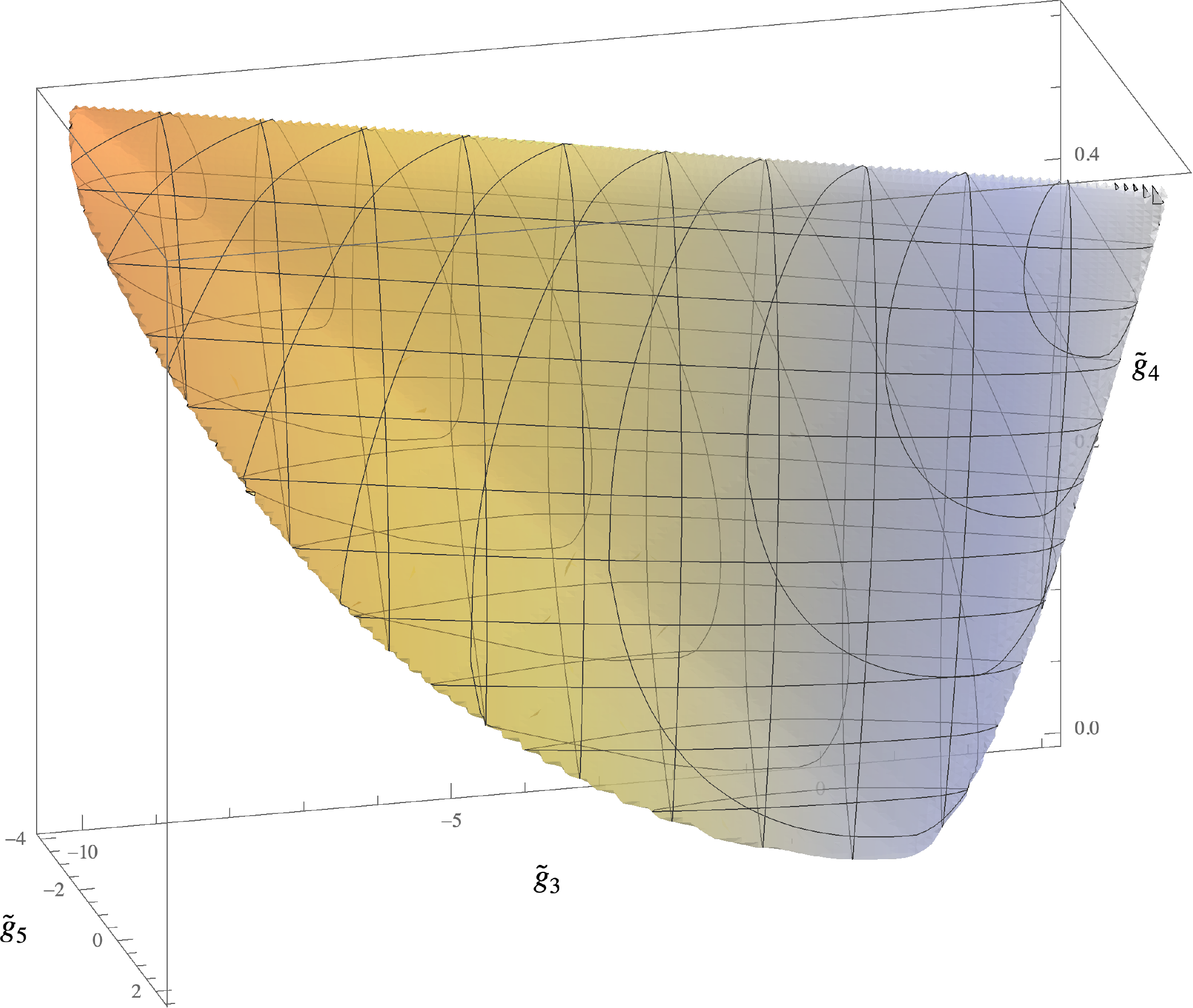}
      \caption{General shape of the volume.}
    \end{subfigure} 
    \begin{subfigure}[t]{0.43\textwidth}
        \centering
  		\includegraphics[width=\linewidth]{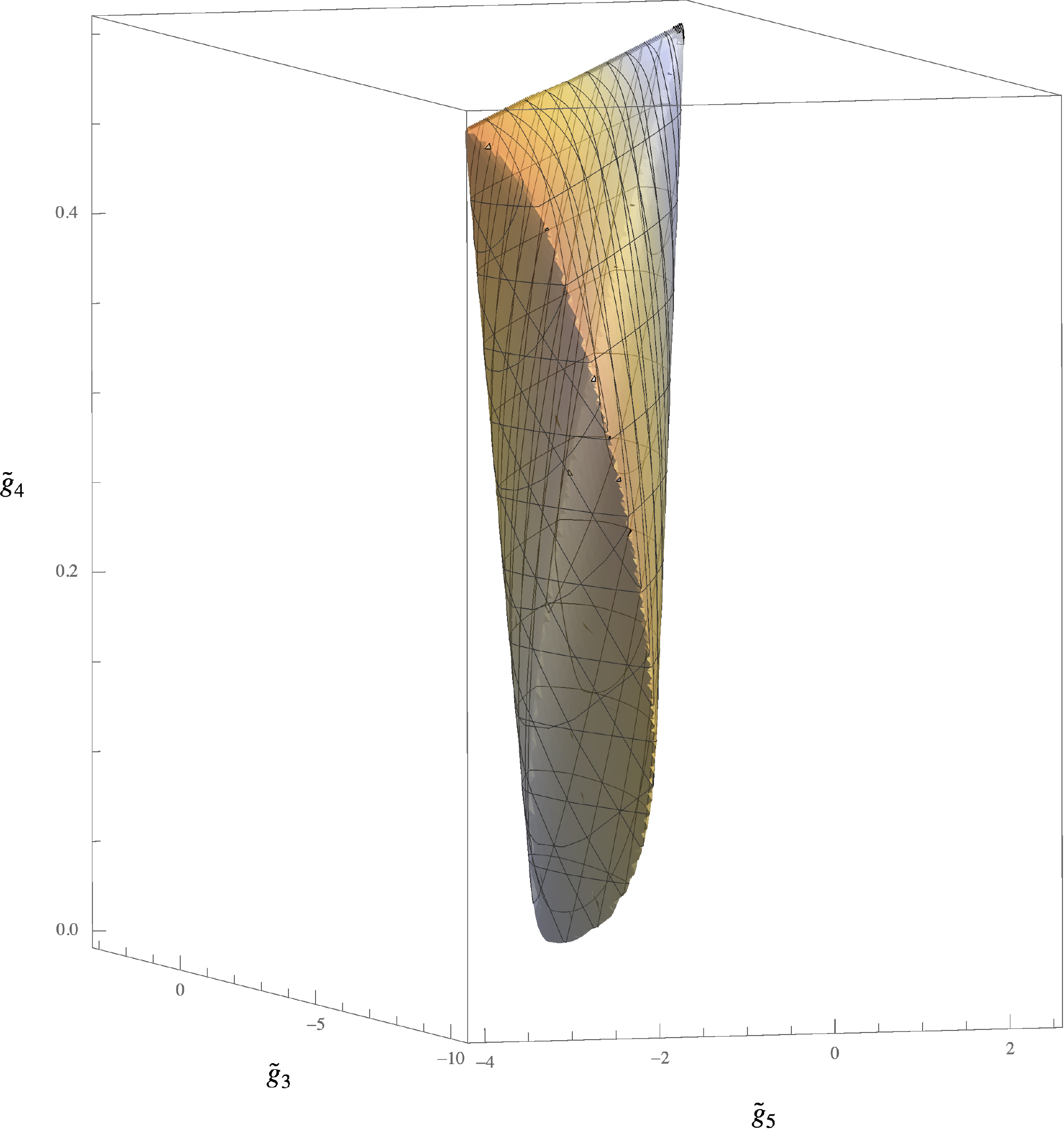}
  		\caption{
		View of the narrow ridge along $\tilde{g}_4 = \tfrac{1}{2}$.
		}
    \end{subfigure}
        \caption{$\tilde{g}_3$ vs. $\tilde{g}_4$ vs. $\tilde{g}_5$ ``tortilla chip".  Numerics were performed at $n=10$ Mandelstam order and $J=0,2,\dots,40$.  The volume is surprisingly narrow, showing a strong $\tilde{g}_3$-$\tilde{g}_5$ correlation.}
        \label{fig:3dcarve}
\end{figure}

Finally, we consider the space of EFT coefficients $(\tilde{g}_3,\tilde{g}_4,\tilde{g}_5)$ using a similar procedure to the $2d$ process.  Once the $2d$ exclusion region is obtained, points in $(\tilde{g}_3, \tilde{g}_4)$ are sampled from the exclusion region and then the optimal upper and lower bounds on $\tilde{g}_5$ with the associated hyperplanes are computed via the constrained optimization approach.  Figure \ref{fig:3dcarve} shows that the $3d$ region is narrow, suggesting that S-matrix positivity is a potent constraint on the EFT.

Finally, the exclusion plot converges as the truncation order $n$ is increased as shown in figures \ref{fig:kinks}.  In particular, the $n=10$ and $n=14$ regions overlap better than the $n=4$ and $n=10$ regions.  Despite doubling the number of functions $n_i(m^2,J)$ in the null constraints $N$ to sharpen bounds between $n=10$ and $n=14$, the numerics show that they generate very similar regions.  Therefore, it is sufficient to consider $n=10$ in order to infer general features.

\section{Analytically ruling-in: the two kinks with $\tilde{g}_4=\frac12$}
\label{sect:analytic}

It is interesting to ask whether the bounds on $(\tilde{g}_3,\tilde{g}_4,\ldots)$ obtained in the preceding section are indeed optimal.
We obtained these bounds using the so-called dual problem, which ``rules-out'' more and more space as one adds null constraints.
This contrasts with the primal problem, whereby the goal is to ``rule-in'' coefficients of $2\to 2$ S-matrices satisfying all the axioms.
When results from both problems agree, the optimal solution is apprehended with complete confidence.

Systematic implementations of the primal problem have been proposed  for generic field theories (without a large gap $M$) \cite{Paulos:2017fhb,Guerrieri:2018uew}. Convergence of the dual and primal problems have also been studied for two-dimensional S-matrices \cite{Cordova:2019lot,Guerrieri:2020kcs}.
We will not attempt to adapt these methods to our problem, but we will study simple special theories which can be ruled in analytically.

The $(\tilde{g}_3,\tilde{g}_4)$ allowed region in fig.~\ref{fig:exclusion} prominently displays two kinks with $\tilde{g}_4=\tfrac12$
(connected with a horizontal segment).
This value is interesting, because the EFT parameter $g_2$ and $g_4$ satisfy the following sum rules:
\be
g_2 = \avg{\frac{1}{m^4}}\ ,\qquad  g_4 = \frac12 \avg{\frac{1}{m^8}}\,,
\ee
where $\avg{\,\cdot \,}$ signifies a positive sum over states with $m\geq M$ and different spin, as previously defined.
The coefficient $g_4=\frac{g_2}{2M^4}$ can only be realized if the high-energy theory contains \emph{only} states
at the single mass $m=M$!

The corresponding amplitudes are therefore rational functions with poles at $s$, $t$ or $u=m^2$.
Furthermore, they must satisfy the boundedness property $\lim_{s\to 0}\frac{\MM(s,t)}{s^2}=0$.
It is easy to see that there are only two crossing-symmetric rational functions with these properties:
\be\begin{aligned}
	\mathcal{M}_{\text{spin-0}} = & \frac{1}{m^2-s} +  \frac{1}{m^2-t} +  \frac{1}{m^2-u}\,, \\
	\mathcal{M}_{stu\text{-pole}} = &  \frac{m^4}{\left(m^2-s\right) \left(m^2-t\right) \left(m^2-u\right)} - \gamma(d) \mathcal{M}_{\text{spin-0}} \label{models}\,.
\end{aligned}\ee
Here, $\gamma(d)$ is a coefficient that will ensure unitarity and positivity of the amplitude in $d$-dimensional space-time.  In particular, we have that 
\be
\gamma(d) \equiv \frac{4}{9} \, _2F_1\left(\tfrac{1}{2},1,\tfrac{d-1}{2},\tfrac{1}{9}\right)\,. \label{gamma d}
\ee

To show this, consider the spectral density of the first amplitude (its imaginary part) is supported entirely on spin-0.
Following the decomposition used in fig.~\ref{fig:decomposition}, we have then subtracted a constant multiple of $\MM_{\text{spin}-0}$
to remove the spin-0 content from $\mathcal{M}_{stu\text{-pole}}$. To find the left kink, we should thus
tune $\gamma(d)$ to make the spectral density supported only on spins $J\geq 2$.

In general, the Gegenbauer polynomials satisfy an orthogonality relation \cite{Correia:2020xtr} that allows one to extract the coefficient:
\begin{align}
f_J(s)=\frac{\mathcal{N}_{d}}{2} \int_{-1}^{1} d z\left(1-z^{2}\right)^{\frac{d-4}{2}} \legP_{J}(z)\mathcal{M}(s,t(z)),
\end{align}
where $z = 1 + \tfrac{2t}{s}$ and $\mathcal{N}_d=\frac{(16 \pi)^{\frac{2-d}{2}}}{\Gamma\left(\frac{d-2}{2}\right)}$
we find
\be
 \rho_J(s)\big|_{stu\text{-pole}} = 2\pi\mathcal{N}_d m^{d-4} \delta(s-m^2) \int_{-1}^1 dz\left(1-z^{2}\right)^{\frac{d-4}{2}} \legP_{J}(z)
 \left(\frac{1}{9-z^2} - \frac{\gamma(d)}{4}\right)\,. \label{rhoJ stu}
\ee


Plugging in $\legP_0(z)=1$ and computing the integral, we find that the spin-0 component vanishes provided we choose:
\be
\gamma(d) = \frac{4}{9} \, _2F_1\left(\tfrac{1}{2},1,\tfrac{d-1}{2},\tfrac{1}{9}\right)\,.
\ee
To verify that the partial waves are positive for $J\geq 2$,
one can use the Froissart-Gribov formula (see eq.~(2.53) of \cite{Correia:2020xtr}) to analytically integrate \eqref{rhoJ stu} in terms of
a single residue at $z=\pm 3$; the integral is proportional to the function defined as $Q_J^{(d)}(3)$ there, which is a positive hypergeometric function for any $d\geq 3$.
We conclude that the amplitude
$\mathcal{M}_{stu\text{-pole}}$, with the value \eqref{gamma d}, satisfies all the axioms of crossing symmetry,
spin-2 Regge convergence, and positivity $\rho_J(s)\geq 0$!
Of course, this does not imply that this amplitude can indeed be realized in some fully-fledged UV complete theory,
only that it cannot be ruled out with our current methods.

\begin{figure}[t]
    \centering
    \includegraphics[width=0.7\textwidth]{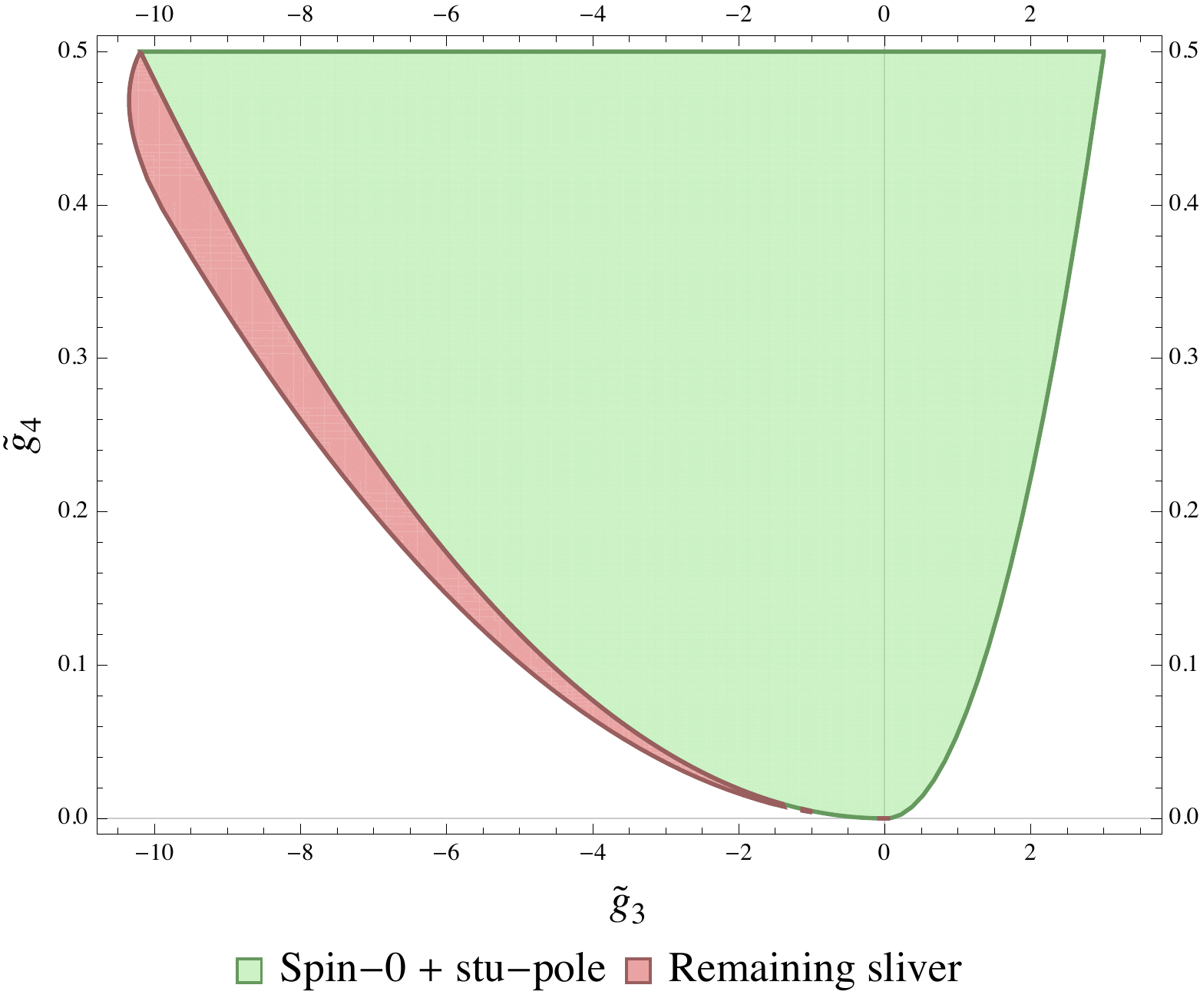}
    \caption{Remaining allowed sliver compared with the positive span of the two analytical models: $\mathcal{M}_{stu\text{-pole}}$ and $\mathcal{M}_{\text{spin-0}}$,
    which lie at the left and right upper kinks, respectively.
    \label{fig:analytic_exclusion}}
\end{figure}

Let us now situate the amplitudes $\mathcal{M}_{\text{spin-0}}$ and $\mathcal{M}_{stu\text{-pole}}$ in the context of our numerical exclusion plots.
By series expanding at small $s,t,u$ and comparing with the low-energy parameterization \ref{eq:gegenbauer_expansion}, it is straightforward to find that
\begin{align}
	\mathcal{M}_{\text{spin-0}}:& \qquad (\tilde{g_3},\tilde{g}_4, \tilde{g}_5)= \left(\frac{3}{\tilde{m}^2},\frac{1}{2 \tilde{m}^4},\frac{5}{2 \tilde{m}^6}\right)\\
	\mathcal{M}_{stu\text{-pole}}:& \qquad (\tilde{g_3},\tilde{g}_4, \tilde{g}_5)=  \left(\frac{2}{\tilde{m}^2} \frac{3 \gamma(d) -1}{2 \gamma(d) -1},\frac{1}{2 \tilde{m}^4},\frac{1}{\tilde{m}^6}\frac{5 \gamma(d) -2}{2 \gamma(d) -1}\right),
\end{align}
where $\tilde{m} \equiv \frac{m}{M}$. 
At $\tilde{m}=1$ and $d=4$, the spin-$0$ model produces the upper-right kink at $(3,\tfrac{1}{2})$,
whereas the $stu$-pole model gives $(\tilde{g_3},\tilde{g}_4)=(6\frac{2\log 2-1}{4\log 2-3},\tfrac12)\approx (-10.19196,0.5)$
which precisely matches with the numerical value in fig.~\ref{fig:kinks}!
We also found agreement with the numerical position of the kink in $d=6$ and $d=10$. As $d\to\infty$, the kink converges to $(-6,-\frac12)$.

The fact that the amplitude $\mathcal{M}_{stu\text{-pole}}$ realizes a negative coefficient for $stu$ (i.e. $g_3<0)$ suggests that it would be impossible
to prove irreversibility of six-dimensional renormalization group flow using only crossing and positivity of $2\to 2$ scattering \cite{Elvang:2012st}.
This contrasts the successful four-dimensional case, in which the $a$-theorem was related to positivity of $g_2$ \cite{Komargodski:2011vj}.

The convex hull of the space of the two amplitudes $\mathcal{M}_{\text{spin-0}}$ and $\mathcal{M}_{stu\text{-pole}}$, for different values of the mass $m$,
generates the region with parabolic boundaries shown in fig.~\ref{fig:analytic_exclusion}.
The parabolic ``underbellies'' are simply the dimensionally-rescaled theories at the kinks.
This region is analytically ``ruled in''.

One notices that a sliver obtained via numerics is not in the span of these simple analytic models.
Since the upper-left arc of this sliver cannot be generated by the convex hull of (possibly rescaled) discrete theories,
we attribute it to a continuous one-parameter family of ``extremal theories'' which terminate at the kink.
It would be interesting to find an analytic expressions for this family.

One may wonder if this family of extremal theories terminates at a second kink;
we did not locate any stable candidate in the numerics. This suggests that the family instead disappears inside its own convex hull.
Naturally this would occur at the point where its slope becomes tangent to the parabola of dimensionally-rescaled theories originating from that point,
$\frac{dg_3}{g_4}=\frac{g_3}{2g_4}$.  In this scenario, only the second derivative of the boundary shape would be discontinuous,
explaining the difficulty in locating it numerically.

We find it remarkable that two simple analytic models almost span the entire region obtained via numerics:
to a good approximation (up to the missing sliver),
\emph{a scalar EFT is compatible with causality and unitarity if it is a positive linear combination
of the models in eq.~\eqref{models}}.

\section{Concluding remarks}
\label{sect:conclusion}

In this work we showed, in the case of a scalar field theory, that
the space of low effective field theories is sharply constrained by positivity of the S-matrix.
Dimensional analysis teaches us to expect a low-energy coupling of mass dimension $n$ to be
suppressed by a factor of $1/M^n$, where $M$ is the mass of new heavy states.
Our main finding is simple: \emph{in any causal and unitary theory, dimensional analysis scaling is a theorem}.

More precisely, for a theory of a single identical real scalar, we
showed that dimensionless ratios of the form $g_k/g_2 M^\#$, where $g_2$ is the coefficient of $\tfrac{1}{2}(\partial\phi)^4$, are bounded above and below by finite constants of order unity.  Our method does not assume that physics above the scale $M$ is weakly coupled, only that it is consistent with unitarity and causality.
The technical assumption is the convergence of double-subtracted dispersion relations in $2\to 2$ scattering.
We then separately bounded $g_2$ itself in eq.~\eqref{g2 upper}.

To our (possibly incomplete) knowledge, this is the first time that two-sided bounds are obtained for interactions that vanish in the forward limit,
such as the six-derivative ``$stu$'' contact interaction $g_3$ (see eqs.~\eqref{eqn:eft_amplitude} and \eqref{eqn:box_bounds}).
The key ingredient was to use ``null constraints'' (for example eq.~\eqref{eqn:null1}):
integrals over the high-energy spectral density which must vanish by crossing symmetry, and which
limit the contribution of higher-spin particles. A systematic procedure to extract optimal bounds was presented.
The precise form of the null constraints are affected by low-energy self-interactions (i.e. loops within the low-energy EFT),
whose effects would be interesting to investigate.
It would be interesting to assess if bounds of this type are closely saturated in specific low-energy processes, for example pion scattering.
In this case, a generalization to non-identical scalars might be necessary.

In this paper we ignored gravity.
A graviton pole would cause the $B_2(t)$ sum rule to diverge in the forward limit, invalidating conclusions from a Taylor series around $t=0$.
However the $X_2(t)$ null constraints in eq.~\eqref{X2} should remain valid for $t<0$ and their implications are worth investigating.
Intuitively, one may anticipate that the graviton pole will somewhat weaken the bounds on scalar scattering \cite{Adams:2006sv,Tokuda:2020mlf,Alberte:2020jsk} (see also \cite{Bellazzini:2019xts}). However, following the general principle that self-consistent spinning S-matrices are harder to come by, interactions involving gravitons are likely to be sharply constrained by similar methods, possibly extending \cite{Camanho:2014apa,Hamada:2018dde,deRham:2018qqo} or addressing conjectures of \cite{Chowdhury:2019kaq}.

For non-gravitational scalar scattering, we found that crossing symmetry does not mix heavy spinning and spinless states.  The allowed $2\to 2$ low-energy S-matrices are positive sums of those two sectors.
These are respectively (almost completely) spanned by the two simple analytic models in eq.~\eqref{models}, which realize theories at kinks.
Perhaps this is a tantalizing hint that the collection of valid EFTs is not so vast after all.

\acknowledgments

We thank Dalimil Maz\`{a}\v{c}, Leonardo Rastelli, David Simmons-Duffin and Anh-Khoi Trinh for useful conversations and Cl\'ement Virally for technical assistance.
Work of VVD is supported by the National Science and Engineering Council of Canada.
Work of SCH is supported by the National Science and Engineering Council of Canada, the Canada Research Chair program,
the Fonds de Recherche du Qu\'ebec - Nature et Technologies, and the Simons Collaboration on the Nonperturbative Bootstrap.  

\appendix

\section{A basis of crossing symmetry constraints}

\label{app:Xn}

In this section we present a complete basis of vanishing heavy averages, which vanish
in any theory with no branch cut below $M^2$ (i.e. when neglecting low-energy loops).
They are organized as functions $X_k(t)$ where $k$ is an even integer representing the number of subtractions.

We first give a pedestrian argument considering the unsubtracted case, $X_0$,
and then make a general argument exploiting crossing symmetry.

As a warm-up, let us first consider a ``superbounded''
theory where $\lim_{s\to\infty}\MM(s,t)\to 0$ for $t<0$, so that unsubtracted dispersion relations converge,
and where the low-energy amplitude is polynomial, corresponding to $g=0$ in \eqref{eqn:eft_amplitude}.
In the superbounded case the first sum rule is $B_0$ from eq.~\eqref{Bks general}:
\be
 B_0: \ g_0 + 2g_2 t^2 +2g_4 t^4 + \ldots  = \avg{\frac{2m^2+t}{m^2+t}\legP\left(1+\frac{2t}{m^2}\right)}
\ee 
where we relabelled the constant term in the amplitude as $g_0=-\lambda$.
Viewed as a function of $t$ this sum rule is marginally useful because it involves infinitely many unknowns on the left-hand-side
and one loses control when $-t$ becomes of order $M^2$.
However, each $g_k$ with $k\geq 2$ can be computed by some other sum rule: 
for example, the $t\to 0$ limit of $B_2$ measures $g_2$, $B_4(0)$ measures $g_4$, etc.
Dividing by an overall $t$ for future convenience, we can package these into an infinite set of null constraints:
\be\begin{aligned}
 0 &= \avg{\frac{2m^2+t}{t(m^2+t)}\legP_J\left(1+\frac{2t}{m^2}\right)} - \sum_{k=0}^{\infty} B_k^{\rm high}(0) t^{2k-1}
\\
  &= \avg{X_0(t; m^2,j)} \quad\mbox{where}\quad
  X_0(t;m^2,J)\equiv \frac{(2m^2+t)\legP_J\left(1+\frac{2t}{m^2}\right)}{t(m^2+t)} - \frac{2m^4\legP_J(1)}{t(m^4-t^2)}\,.
\label{X0}
\end{aligned}\ee
$X_0(t)$ defines a sensible sum rule for any $0<-t<M^2$, and Taylor-expanding around $t=0$ gives an infinite
number of averages which vanish by crossing symmetry (if unsubtracted sum rules converge).

Repeating the same manipulations for subtracted sum rules, we find that to
cancel the cancel the generic power of $t$ in $B_2(t)$ one must also both the $B_k(0)$ and its first derivative around $t=0$;
there is then a unique solution: $\langle X_2(t;m^2,J)\rangle=0$ where
\be \label{X2}
 X_2(t;m^2,J) = \frac{(2m^2+t)\left(\legP_J\left(1+\frac{2t}{m^2}\right)-\legP_J(1)\right)}{m^2t^2(m^2+t)^2} -
 \frac{4 \legP_J^\prime(1)}{t m^2(m^4-t^2)}\,.
\ee
For any $0<-t<M^2$, this enjoys the high-energy convergence as a double-subtracted dispersion relation,
whence the subscript. Setting $t=0$ this reproduces the $n_4$ sum rule in eq.~\eqref{null constraints}.
Finally, we record explicitly the four-subtracted version obtained with the same method:
\be
 X_4(t;m^2,J) = \frac{(2m^2+t)\left(\legP_J\left(1+\frac{2t}{m^2}\right)-\legP_J(1)\right)}{m^4 t^3(m^2+t)^3}
-\frac{2(2m^4-3t(m^2+t))\legP_J^\prime(1)}{m^8t^2(m^2-t)(m^2+t)^2} -\frac{4 \legP_J^{\prime\prime}(1)}{m^8 t(m^4-t^2)}\,.
\label{X4}
\ee

\subsection{Derivation using dispersion relations in three channels}

Let us now present the general case along with a direct derivation.
The idea is to combine dispersion relations in all three channels.
Consider the following identity:
\be
 \label{Xk neat}
X_k(t'):\ 0=\left[ \oint_{C_s}+ \oint_{C_t}+\oint_{C_u}\right] \frac{ds\wedge dt}{(2\pi i)^2s t (t-t')} \frac{\MM(s,t)}{[st(s+t)]^{k/2}}\,,
\ee
where each contour is a product of circles: $C_s= {\rm Circle}_{s\sim 0} \wedge {\rm Circle}_{t\to \infty}$,
$C_t= {\rm Circle}_{t\sim 0,t'} \wedge {\rm Circle}_{u\to \infty}$ and $C_u= {\rm Circle}_{u\sim 0} \wedge {\rm Circle}_{s\to \infty}$.
The first contour implements a fixed-$s$ sum rule, which we then evaluate at $s\to 0$, and similar for the others, but note that
the fixed-$t$ relation is evaluated at both $t=0$ and $t=t'$.
All the integrals vanish (for $k\geq 2$) due to the vanishing at large-$t$ of $\MM(s,t)/t^2$.
(The contours around $0$ are a bit dangerous, since the integrals around say large-$t$ converges only when $s<0$;
this should not be a problem since (ignoring EFT loops) one can interpret the residue as extracting the coefficient of $1/s$
in the Laurent expansion as $s\to 0^-$.)
The trick is then to deform the arcs at infinity to pick high-energy cuts and low-energy poles, as in fig~\ref{fig:contour_deformation} of the main text.

What is nice with the above contour is that all double-residues cancel because of antisymmetry of the contours.
For the double residue at $(s,t)=(0,t')$ the cancellation simply follows from mismatching orientations,
since the contours $C_s$ and $C_t$ compute the same double-residue in opposite orders:
\be
 0 = {\rm Circle}_{s\sim 0} \wedge {\rm Circle}_{t\sim t'} +{\rm Circle}_{t\sim t'}\wedge {\rm Circle}_{s\sim 0}\,.
\ee
Alternatively, in practice,
one may think of the integral $\frac{1}{2\pi i}\int_{{\rm Circle}_{s\sim 0}}$ as just a residue, defined by plucking $\frac{ds}{s}\wedge$ from the left of the differential form it multiplies,
and the integral vanishes because the thus-defined residue operation is antisymmetric.
The nested poles at $(s,t)=(0,0)$ require more care (one has to perform a blow-up) since all three contours contribute,
but one still finds a perfect cancellation, as one may verify explicitly:
\be \label{Jacobi}
 \left(\ \Res\limits_{t=0}\Res\limits_{s=0} \ +\
  \Res\limits_{s=0}\Res\limits_{t=0} \ +\  \Res\limits_{t=0}\Res\limits_{s=-t}\right) \frac{{\rm Poly}(s,t)}{(s t(s+t))^n}ds\wedge dt= 0
\ee
for any polynomial numerator. This is curiously reminiscent of the Jacobi identity.

The upshot is that the integral \eqref{Xk neat} is orthogonal to \emph{any} tree-level low-energy EFT amplitude (whether or not it is $s$-$t$-$u$
symmetric), it gives purely a constraint on high-energy cuts.
Similar constraints would follow for any choice of (rational) denominator.
Each term in eq.~\eqref{Xk neat} contributes cuts in two channels, for example the $C_s$ term contributes
\be
 \avg{ \Res\limits_{s=0} \left(\frac{1}{m^2(t'-m^2)}+ \frac{1}{(m^2+s)(m^2+s+t')}\right) \frac{m^2\legP_J\left(1+\frac{2s}{m^2}\right)}{s(sm^2(m^2+s))^{k/2}} }
\ee
where the two terms in the parenthesis come from the $t$ and $u$-channel heavy cuts, respectively.
If we did not assume $s{-}t{-}u$ symmetry, we would get a relation between the three spectral densities.
Here we record only the simplified result assuming that all the spectral densities are the same: $0=\langle X_k(t;m^2,J)\rangle$
where
\be\begin{aligned}
X_k(t;m^2,J)=& \frac{2m^2+t}{t(m^2+t)}\frac{\legP_J\left(1+\frac{2t}{m^2}\right)}{(tm^2(m^2+t))^{k/2}}
 \\ &\ -\Res\limits_{s=0} \left[\frac{(2m^2+s)(m^2-s)(m^2+2s)}{s(t-s)(m^2+s)(m^2-t)(m^2+s+t)}\frac{\legP_J\left(1+\frac{2s}{m^2}\right)}{(sm^2(m^2+s))^{k/2}}\right]\,.
\end{aligned}\ee
This is the main result of this appendix. For any even $k\geq 2$ and $-M^2<t<0$ we expect these to be physical sum rules
(i.e. convergent in theories where double-subtracted dispersion relations converge) orthogonal to any tree-level low-energy amplitude.
For $k=0,2,4$ this reproduces the formulas quoted in eqs.~\eqref{X0}-\eqref{X4}.

Expanding at small $t$, the $X_k(t)$ sum rules admit regular Taylor series, which reproduce precisely the sum rules recorded in eq.~\eqref{null constraints}.
Namely, the coefficient of $t^n$ in $X_k$ has degree $\frac{3k}{2}+n+1$ in $1/m^2$, so the first case is $X_2(0) \sim g_4$.
The first time one gets two $X$ sum rules is at weight 7, where $X_2'''(0)$ and $X_4(0)$ span $g_7$ and $g_7'$.
The number of $X$ sum rules per degree increases every 3 degree because of the $1/(stu)^{k/2}$ factor in eq.~\eqref{Xk neat}.
The number of $X$ sum rules thus agrees precisely with the counting below eq.~\eqref{null constraints}.
We conclude that the $X_k(t)$ sum rules are a complete basis of sum rules orthogonal to tree-level EFTs!
As mentioned below eq.~\eqref{scheme indep}, when EFT loops are included these sum rules may average to nonzero
but computable quantities.

\begin{table}[t]
\centering
{\renewcommand{\arraystretch}{1.7}}
\begin{tabular}{|c|c|c|}
\hline
\textbf{EFT coefficient} & \textbf{Lower bound} & \textbf{Upper bound} \\ \hline
$\tilde{g}_{3}$          & -10.346         & 3                    \\ \hline
$\tilde{g}_{4}$          & 0                    & 0.5                  \\ \hline
$\tilde{g}_{5}$          & -4.096         & 2.5                  \\ \hline
$\tilde{g}_{6}$    & 0                    & 0.25                 \\ \hline
$\tilde{g}_{6}^\prime$    & -12.83         & 3                    \\ \hline
$\tilde{g}_{7}$          & -1.548         & 1.75                 \\ \hline
$\tilde{g}_{8}$    & 0                    & 0.125                \\ \hline
$\tilde{g}_{8}^\prime$    & -10.03         & 4                    \\ \hline
$\tilde{g}_{9}$    & -0.524        & 1.125                \\ \hline
$\tilde{g}_{9}^\prime$    & -13.60         & 3                    \\ \hline
$\tilde{g}_{10}$   & 0                    & 0.0625               \\ \hline
$\tilde{g}_{10}^\prime$   & -6.32         & 3.75                 \\ \hline
\end{tabular}
\caption{Bounds on coefficients $\tilde{g}_k^{(p)}=g_k^{(p)}M^{2k-4}/g_2$ for $d=4$ spacetime dimension, where $g_k^{(p)}$
refers to the coefficient
of $(s^2+t^2+u^2)^{\frac{k-3(2p+\delta_{k,\rm odd})}{2}} (stu)^{2p+\delta_{k,\rm odd}}$,
which has degree $k$ in Mandelstam invariants and contains $2p$ powers of $stu$ more than the minimum at that degree.
The upper bounds are all simple rational numbers realized by the $\mathcal{M}_{\text{spin-0}}$ model.  The values (except for $\tilde{g}_3$)
were calculated at order $n=10$, which corresponds to the number of null constraints of $\text{dim} \,N = 12$. 
\label{tab:4d}
}
\end{table}

\section{Bounds on operators up to order $s^{10}$}
\label{app:operator}

In table \ref{tab:4d} we record numerical bounds on various EFT coefficients in four spacetime dimensions.

\bibliographystyle{JHEP}
\bibliography{ref}{}

\providecommand{\href}[2]{#2}\begingroup\raggedright\begin{thebibliography}{10}

\bibitem{Adams:2006sv}
A.~Adams, N.~Arkani-Hamed, S.~Dubovsky, A.~Nicolis and R.~Rattazzi,
  \emph{{Causality, analyticity and an IR obstruction to UV completion}},
  \href{https://doi.org/10.1088/1126-6708/2006/10/014}{\emph{JHEP} {\bfseries
  10} (2006) 014} [\href{https://arxiv.org/abs/hep-th/0602178}{{\ttfamily
  hep-th/0602178}}].

\bibitem{Martin:1969ina}
A.~Martin, \emph{{Scattering Theory: Unitarity, Analyticity and Crossing}},
  vol.~3 (1969), \href{https://doi.org/10.1007/BFb0101043}{10.1007/BFb0101043}.

\bibitem{Roy:1971tc}
S.M.~Roy, \emph{{Exact integral equation for pion pion scattering involving
  only physical region partial waves}},
  \href{https://doi.org/10.1016/0370-2693(71)90724-6}{\emph{Phys. Lett. B}
  {\bfseries 36} (1971) 353}.

\bibitem{Colangelo:2001df}
G.~Colangelo, J.~Gasser and H.~Leutwyler, \emph{{$\pi \pi$ scattering}},
  \href{https://doi.org/10.1016/S0550-3213(01)00147-X}{\emph{Nucl. Phys. B}
  {\bfseries 603} (2001) 125}
  [\href{https://arxiv.org/abs/hep-ph/0103088}{{\ttfamily hep-ph/0103088}}].

\bibitem{Caprini:2003ta}
I.~Caprini, G.~Colangelo, J.~Gasser and H.~Leutwyler, \emph{{On the precision
  of the theoretical predictions for pi pi scattering}},
  \href{https://doi.org/10.1103/PhysRevD.68.074006}{\emph{Phys. Rev. D}
  {\bfseries 68} (2003) 074006}
  [\href{https://arxiv.org/abs/hep-ph/0306122}{{\ttfamily hep-ph/0306122}}].

\bibitem{Pham:1985cr}
T.N.~Pham and T.N.~Truong, \emph{{Evaluation of the Derivative Quartic Terms of
  the Meson Chiral Lagrangian From Forward Dispersion Relation}},
  \href{https://doi.org/10.1103/PhysRevD.31.3027}{\emph{Phys. Rev. D}
  {\bfseries 31} (1985) 3027}.

\bibitem{Ananthanarayan:1994hf}
B.~Ananthanarayan, D.~Toublan and G.~Wanders, \emph{{Consistency of the chiral
  pion pion scattering amplitudes with axiomatic constraints}},
  \href{https://doi.org/10.1103/PhysRevD.51.1093}{\emph{Phys. Rev. D}
  {\bfseries 51} (1995) 1093}
  [\href{https://arxiv.org/abs/hep-ph/9410302}{{\ttfamily hep-ph/9410302}}].

\bibitem{Camanho:2014apa}
X.O.~Camanho, J.D.~Edelstein, J.~Maldacena and A.~Zhiboedov, \emph{{Causality
  Constraints on Corrections to the Graviton Three-Point Coupling}},
  \href{https://doi.org/10.1007/JHEP02(2016)020}{\emph{JHEP} {\bfseries 02}
  (2016) 020} [\href{https://arxiv.org/abs/1407.5597}{{\ttfamily 1407.5597}}].

\bibitem{Afkhami-Jeddi:2016ntf}
N.~Afkhami-Jeddi, T.~Hartman, S.~Kundu and A.~Tajdini, \emph{{Einstein gravity
  3-point functions from conformal field theory}},
  \href{https://doi.org/10.1007/JHEP12(2017)049}{\emph{JHEP} {\bfseries 12}
  (2017) 049} [\href{https://arxiv.org/abs/1610.09378}{{\ttfamily
  1610.09378}}].

\bibitem{Cheung:2016yqr}
C.~Cheung and G.N.~Remmen, \emph{{Positive Signs in Massive Gravity}},
  \href{https://doi.org/10.1007/JHEP04(2016)002}{\emph{JHEP} {\bfseries 04}
  (2016) 002} [\href{https://arxiv.org/abs/1601.04068}{{\ttfamily
  1601.04068}}].

\bibitem{talks}
N.~Arkani-Hamed and Y.-T.~Huang, \emph{Positive geometry of effective field theory,} lectures at the CERN winter school on supergravity, strings and gauge theory (2019);
\emph{New positivity bounds from the EFT hedron,} talk at the 24th rencontres
  Itzykson of the IPHT of CEA-Saclay (2019).

\bibitem{deRham:2017avq}
C.~de~Rham, S.~Melville, A.J.~Tolley and S.-Y.~Zhou, \emph{{Positivity bounds
  for scalar field theories}},
  \href{https://doi.org/10.1103/PhysRevD.96.081702}{\emph{Phys. Rev. D}
  {\bfseries 96} (2017) 081702}
  [\href{https://arxiv.org/abs/1702.06134}{{\ttfamily 1702.06134}}].

\bibitem{Bellazzini:2020cot}
B.~Bellazzini, J.~Elias~Mir\'o, R.~Rattazzi, M.~Riembau and F.~Riva,
  \emph{{Positive Moments for Scattering Amplitudes}},
  \href{https://arxiv.org/abs/2011.00037}{{\ttfamily 2011.00037}}.

\bibitem{Tolley:2020gtv}
A.J.~Tolley, Z.-Y.~Wang and S.-Y.~Zhou, \emph{{New positivity bounds from full
  crossing symmetry}},  \href{https://arxiv.org/abs/2011.02400}{{\ttfamily
  2011.02400}}.

\bibitem{Brivio:2017vri}
I.~Brivio and M.~Trott, \emph{{The Standard Model as an Effective Field
  Theory}}, \href{https://doi.org/10.1016/j.physrep.2018.11.002}{\emph{Phys.
  Rept.} {\bfseries 793} (2019) 1}
  [\href{https://arxiv.org/abs/1706.08945}{{\ttfamily 1706.08945}}].

\bibitem{Correia:2020xtr}
M.~Correia, A.~Sever and A.~Zhiboedov, \emph{{An Analytical Toolkit for the
  S-matrix Bootstrap}},  \href{https://arxiv.org/abs/2006.08221}{{\ttfamily
  2006.08221}}.

\bibitem{GellMann:1954db}
M.~Gell-Mann, M.~Goldberger and W.E.~Thirring, \emph{{Use of causality
  conditions in quantum theory}},
  \href{https://doi.org/10.1103/PhysRev.95.1612}{\emph{Phys. Rev.} {\bfseries
  95} (1954) 1612}.

\bibitem{Bros:1965kbd}
J.~Bros, H.~Epstein and V.~Glaser, \emph{{A proof of the crossing property for
  two-particle amplitudes in general quantum field theory}},
  \href{https://doi.org/10.1007/BF01646307}{\emph{Commun. Math. Phys.}
  {\bfseries 1} (1965) 240}.

\bibitem{SCH:bootstrap2020}
S.~Caron-Huot,
  \emph{{\href{https://projects.iq.harvard.edu/bootstrap2020/videos}{Lorentzian
  and Analytic Bootstrap Lecture 3}}}, {\emph{2020 Bootstrap School} (2020) }.

\bibitem{Jin:1964zza}
Y.~Jin and A.~Martin, \emph{{Number of Subtractions in Fixed-Transfer
  Dispersion Relations}},
  \href{https://doi.org/10.1103/PhysRev.135.B1375}{\emph{Phys. Rev.} {\bfseries
  135} (1964) B1375}.

\bibitem{Martin:1965jj}
A.~Martin, \emph{{Extension of the axiomatic analyticity domain of scattering
  amplitudes by unitarity. 1.}},
  \href{https://doi.org/10.1007/BF02720568}{\emph{Nuovo Cim. A} {\bfseries 42}
  (1965) 930}.

\bibitem{Eden:1966dnq}
R.J.~Eden, P.V.~Landshoff, D.I.~Olive and J.C.~Polkinghorne, \emph{{The
  analytic S-matrix}}, Cambridge Univ. Press, Cambridge (1966).

\bibitem{Caron-Huot:2020adz}
S.~Caron-Huot, D.~Mazac, L.~Rastelli and D.~Simmons-Duffin, \emph{{Dispersive
  CFT Sum Rules}},  \href{https://arxiv.org/abs/2008.04931}{{\ttfamily
  2008.04931}}.

\bibitem{Penedones:2019tng}
J.~Penedones, J.A.~Silva and A.~Zhiboedov, \emph{{Nonperturbative Mellin
  Amplitudes: Existence, Properties, Applications}},
  \href{https://doi.org/10.1007/JHEP08(2020)031}{\emph{JHEP} {\bfseries 08}
  (2020) 031} [\href{https://arxiv.org/abs/1912.11100}{{\ttfamily
  1912.11100}}].

\bibitem{Elvang:2012st}
H.~Elvang, D.Z.~Freedman, L.-Y.~Hung, M.~Kiermaier, R.C.~Myers and S.~Theisen,
  \emph{{On renormalization group flows and the a-theorem in 6d}},
  \href{https://doi.org/10.1007/JHEP10(2012)011}{\emph{JHEP} {\bfseries 10}
  (2012) 011} [\href{https://arxiv.org/abs/1205.3994}{{\ttfamily 1205.3994}}].

\bibitem{Simmons-Duffin:2015qma}
D.~Simmons-Duffin, \emph{{A Semidefinite Program Solver for the Conformal
  Bootstrap}}, \href{https://doi.org/10.1007/JHEP06(2015)174}{\emph{JHEP}
  {\bfseries 06} (2015) 174}
  [\href{https://arxiv.org/abs/1502.02033}{{\ttfamily 1502.02033}}].

\bibitem{Chester:2019ifh}
S.M.~Chester, W.~Landry, J.~Liu, D.~Poland, D.~Simmons-Duffin, N.~Su et~al.,
  \emph{{Carving out OPE space and precise $O(2)$ model critical exponents}},
  \href{https://doi.org/10.1007/JHEP06(2020)142}{\emph{JHEP} {\bfseries 06}
  (2020) 142} [\href{https://arxiv.org/abs/1912.03324}{{\ttfamily
  1912.03324}}].

\bibitem{Cordova:2019lot}
L.~Cordova, Y.~He, M.~Kruczenski and P.~Vieira, \emph{{The O(N) S-matrix
  Monolith}}, \href{https://doi.org/10.1007/JHEP04(2020)142}{\emph{JHEP}
  {\bfseries 04} (2020) 142}
  [\href{https://arxiv.org/abs/1909.06495}{{\ttfamily 1909.06495}}].

\bibitem{Paulos:2017fhb}
M.F.~Paulos, J.~Penedones, J.~Toledo, B.C.~van Rees and P.~Vieira, \emph{{The
  S-matrix bootstrap. Part III: higher dimensional amplitudes}},
  \href{https://doi.org/10.1007/JHEP12(2019)040}{\emph{JHEP} {\bfseries 12}
  (2019) 040} [\href{https://arxiv.org/abs/1708.06765}{{\ttfamily
  1708.06765}}].

\bibitem{Guerrieri:2018uew}
A.L.~Guerrieri, J.~Penedones and P.~Vieira, \emph{{Bootstrapping QCD Using Pion
  Scattering Amplitudes}},
  \href{https://doi.org/10.1103/PhysRevLett.122.241604}{\emph{Phys. Rev. Lett.}
  {\bfseries 122} (2019) 241604}
  [\href{https://arxiv.org/abs/1810.12849}{{\ttfamily 1810.12849}}].

\bibitem{Guerrieri:2020kcs}
A.L.~Guerrieri, A.~Homrich and P.~Vieira, \emph{{Dual S-matrix Bootstrap I: 2D
  Theory}},  \href{https://arxiv.org/abs/2008.02770}{{\ttfamily 2008.02770}}.

\bibitem{Komargodski:2011vj}
Z.~Komargodski and A.~Schwimmer, \emph{{On Renormalization Group Flows in Four
  Dimensions}}, \href{https://doi.org/10.1007/JHEP12(2011)099}{\emph{JHEP}
  {\bfseries 12} (2011) 099} [\href{https://arxiv.org/abs/1107.3987}{{\ttfamily
  1107.3987}}].

\bibitem{Tokuda:2020mlf}
J.~Tokuda, K.~Aoki and S.~Hirano, \emph{{Gravitational positivity bounds}},
  \href{https://arxiv.org/abs/2007.15009}{{\ttfamily 2007.15009}}.

\bibitem{Alberte:2020jsk}
L.~Alberte, C.~de~Rham, S.~Jaitly and A.J.~Tolley, \emph{{Positivity Bounds and
  the Massless Spin-2 Pole}},
  \href{https://arxiv.org/abs/2007.12667}{{\ttfamily 2007.12667}}.

\bibitem{Bellazzini:2019xts}
B.~Bellazzini, M.~Lewandowski and J.~Serra, \emph{{Positivity of Amplitudes,
  Weak Gravity Conjecture, and Modified Gravity}},
  \href{https://doi.org/10.1103/PhysRevLett.123.251103}{\emph{Phys. Rev. Lett.}
  {\bfseries 123} (2019) 251103}
  [\href{https://arxiv.org/abs/1902.03250}{{\ttfamily 1902.03250}}].

\bibitem{Hamada:2018dde}
Y.~Hamada, T.~Noumi and G.~Shiu, \emph{{Weak Gravity Conjecture from Unitarity
  and Causality}},
  \href{https://doi.org/10.1103/PhysRevLett.123.051601}{\emph{Phys. Rev. Lett.}
  {\bfseries 123} (2019) 051601}
  [\href{https://arxiv.org/abs/1810.03637}{{\ttfamily 1810.03637}}].

\bibitem{deRham:2018qqo}
C.~de~Rham, S.~Melville, A.J.~Tolley and S.-Y.~Zhou, \emph{{Positivity Bounds
  for Massive Spin-1 and Spin-2 Fields}},
  \href{https://doi.org/10.1007/JHEP03(2019)182}{\emph{JHEP} {\bfseries 03}
  (2019) 182} [\href{https://arxiv.org/abs/1804.10624}{{\ttfamily
  1804.10624}}].

\bibitem{Chowdhury:2019kaq}
S.D.~Chowdhury, A.~Gadde, T.~Gopalka, I.~Halder, L.~Janagal and S.~Minwalla,
  \emph{{Classifying and constraining local four photon and four graviton
  S-matrices}}, \href{https://doi.org/10.1007/JHEP02(2020)114}{\emph{JHEP}
  {\bfseries 02} (2020) 114}
  [\href{https://arxiv.org/abs/1910.14392}{{\ttfamily 1910.14392}}].

\end{thebibliography}\endgroup

\end{document}